\documentclass{tcibook}
\usepackage{fancyhea}
\usepackage{work}
\usepackage{bm} 
\usepackage{graphicx}
\usepackage{multirow}
\usepackage{lineno}
\usepackage{threeparttable}
\usepackage[normalem]{ulem}
\usepackage{color}
\usepackage{tcolorbox}
\usepackage[colorlinks = true,
            linkcolor = blue,
            urlcolor  = blue,
            citecolor = blue,
            anchorcolor = blue]{hyperref}

\def\beq{\begin{equation}}
\def\eeq#1{\label{#1}\end{equation}}
\def\eeqn{\end{equation}}

\newenvironment{Eqnarray}%
   {\arraycolsep 0.14em\begin{eqnarray}}{\end{eqnarray}}
\def\beqa{\begin{Eqnarray}}
\def\eeqa#1{\label{#1}\end{Eqnarray}}
\def\eeqan{\end{Eqnarray}}

\let\bar=\overbar

\def\lsim{\mathrel{\raise.3ex\hbox{$<$\kern-.75em\lower1ex\hbox{$\sim$}}}}
\def\gsim{\mathrel{\raise.3ex\hbox{$>$\kern-.75em\lower1ex\hbox{$\sim$}}}}

\def\del{\partial}
\def\Dslash{\not{\hbox{\kern-4pt $D$}}}
\def\dslash{\not{\hbox{\kern-2pt $\del$}}}
\def\pslash{\not{\hbox{\kern-2pt $p$}}}
\def\ETmiss{\not{\hbox{\kern-4pt $E$}}_T}

\def\Dlr{\mathrel{\raise1.5ex\hbox{$\leftrightarrow$\kern-1em\lower1.5ex\hbox{$D$}}}}

\def\MSB{{\bar{M \kern -2pt S}}}
\def\msb{{\bar{\scriptsize M \kern -1pt S}}}

\def\drb{{\bar{\scriptsize D \kern -1pt R}}}

\newcommand{\customfootnotetext}[2]{{
  \renewcommand{\thefootnote}{#1}
  \footnotetext[0]{#2}}}
  
\usepackage{scrextend}

\begin{document}

\parindent=0pt
\parskip=8pt
\setlength{\evensidemargin}{0pt}
\setlength{\oddsidemargin}{0pt}
\setlength{\marginparsep}{0.0in}
\setlength{\marginparwidth}{0.0in}
\marginparpush=0pt

\renewcommand{\chapname}{chap:intro_}
\renewcommand{\chapterdir}{.}
\renewcommand{\arraystretch}{1.25}
\addtolength{\arraycolsep}{-3pt}

\thispagestyle{empty}
\begin{centering}
\mbox{\null}
\rightline{\begin{tabular}{l}
 \end{tabular}}
\vfill

{\Huge\bf The Future of High Energy Physics}\\\vspace{2mm}
{\Huge\bf Software and Computing}

\vskip 0.2in

{\LARGE \bf Report of the 2021  US  Community Study  \\
     on the Future of Particle Physics

                  \smallskip

       \textit{organized by the  APS Division of Particles and Fields}}

\vfill

\vskip 0.2in

\vspace{-12mm}
\begin{center}
    \includegraphics[width=0.8\textwidth]{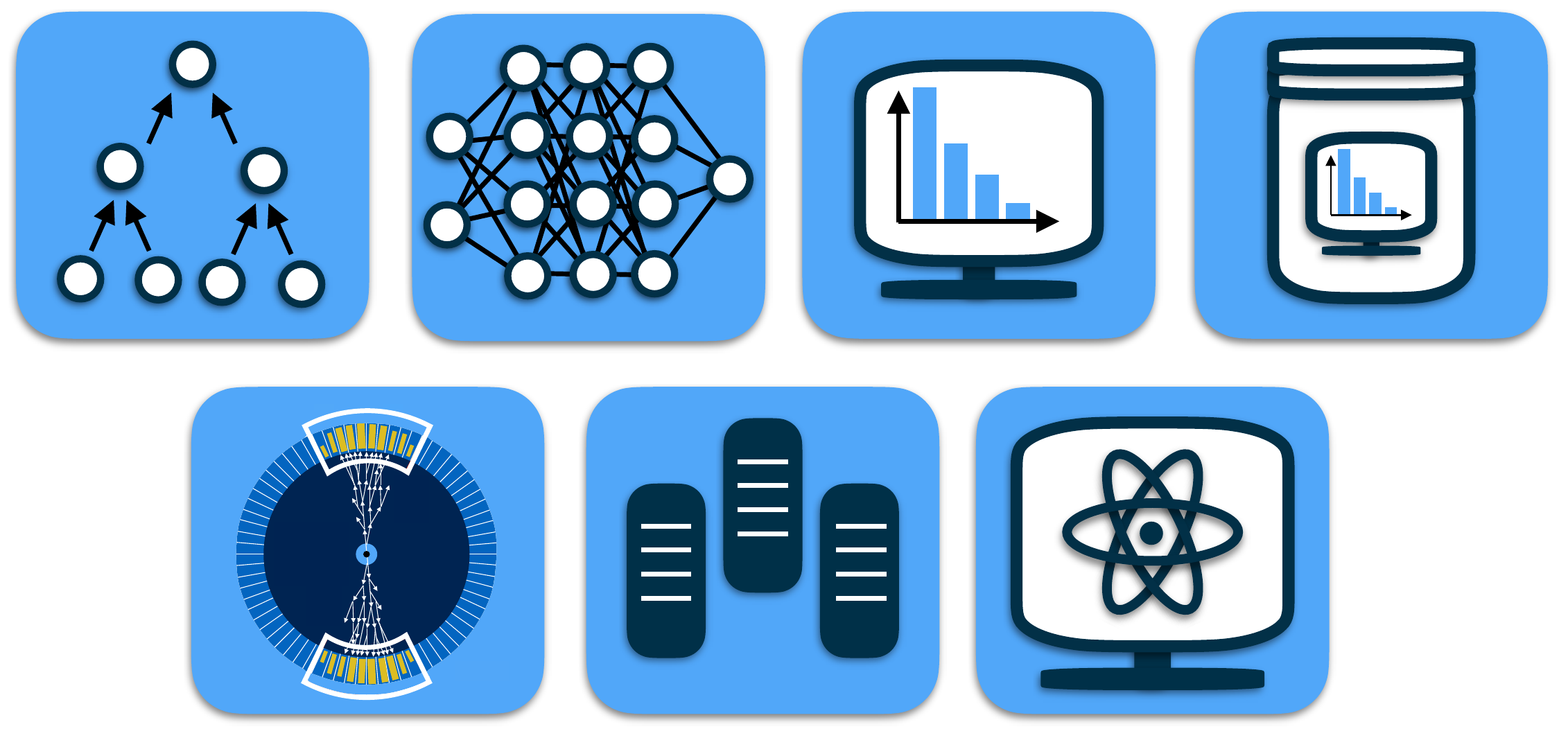}
\end{center}
\vspace{4mm}

\vspace{-4mm}
\begin{center}
v2.1; updated: \today
\end{center}

{\large {\bf Conveners:} V.~Daniel~Elvira\textsuperscript{*}, Steven Gottlieb, Oliver~Gutsche\textsuperscript{\dag}, and Benjamin~Nachman}\\
\vspace{1cm}

{\large {\bf Topical Group Conveners:}
S.~Bailey\textsuperscript{\dag}, W.~Bhimji, P.~Boyle, G.~Cerati, M.~Carrasco~Kind, K.~Cranmer,  G.~Davies, V.~D.~Elvira\textsuperscript{\dag}, 
R.~Gardner\textsuperscript{\dag}, K.~Heitmann, M.~Hildreth\textsuperscript{\#}, W.~Hopkins, T.~Humble, M.~Lin\textsuperscript{*}, P.~Onyisi, J.~Qiang, K.~Pedro\textsuperscript{\dag}, G.~Perdue, A.~Roberts, M.~Savage, P.~Shanahan, K.~Terao, D.~Whiteson, and F.~Wuerthwein} \\  

\customfootnotetext{\dag}{Before Fall '21}
\customfootnotetext{*}{After Fall '21}
\customfootnotetext{\#}{Before Fall '20}

\vfill

\end{centering}

\newpage

\begin{center}
{\LARGE \bf Executive Summary}
\end{center}

\vspace{-3mm}

Software and Computing (S\&C) are essential to all High Energy Physics (HEP) experiments and many theoretical studies.  The size and complexity of S\&C are now commensurate with that of experimental instruments, playing a critical role in experimental design, data acquisition/instrumental control, reconstruction, and analysis.  Furthermore, S\&C often plays a leading role in driving the precision of theoretical calculations and simulations. %
Within this central role in HEP, S\&C has been immensely successful over the last decade.  Every experimental result and many theoretical insights were only possible due to advances in S\&C. We have developed innovative solutions to unique HEP S\&C challenges and when possible, we have embraced community standards in software packages, versioning tools, etc.  Grass-roots initiatives such as the HEP Software Foundation (HSF)~\cite{HEPSoftwareFoundation:2017ggl} have unified parts of the community in these efforts.  A number of successful cross-cutting S\&C research centers and institutes have also emerged to enhance HEP science.

Additionally, limited computing resources have gone much further than anticipated due to methodological innovation.  Part of these advances are driven by increased use of computing hardware heterogeneity and specialization, as well as increased use of high-performance computing facilities.  HEP is a significant user of national supercomputing facilities through large and small~\cite{FASER:2022yqp} particle physics and cosmology experiment and simulation workflows, other areas of theory (predominately lattice field theory), and accelerator modeling.  Significant progress has been made within these communities to adapt software applications for effective use of hardware accelerators, and in preparation for future exascale computing resources. Examples of funded programs in this area include the DOE Exascale Computing Project (ECP), Scientific Discovery through Advanced Computing (SciDAC), the Center for Computational Excellence (CCE), Computational HEP more generally, and the NSF Institute for Research and Innovation in Software for HEP (IRIS-HEP).

Furthermore, the deep learning revolution that started in the last decade is already having a widespread impact on all aspects of HEP.  Quantum computing has emerged as a technology that may enable the study of various quantum systems that are beyond the capabilities of classical tools.  Neither of these topics was considered in the last Snowmass planning process from 2013 and, in both areas, HEP scientists are making contributions well beyond our field.

The goal of Snowmass is to provide input for the Particle Physics Project Prioritization Panel (P5) with a ten year timescale. While S\&C is clearly an enabler of the HEP science drivers, it is not managed like a `project' as in the case of facilities, experiments, and surveys.  S\&C is no less important, often transcends traditional boundaries, and changes on a much faster timescale than Snowmass processes.  For this reason, we have identified one central recommendation for the 2021 Snowmass:
 
\begin{tcolorbox}

 We recommend the creation of a standing  {\color{blue}\textbf{Coordinating Panel for Software and Computing (CPSC)}} under DPF, mirroring the panel for advanced detectors (\href{https://cpad-dpf.org}{CPAD}) established in 2012. \\

\vspace{-2mm}
\begin{addmargin}[1em]{2em}
\textit{Purpose: 
Promote, coordinate, and assist the HEP community on Software and Computing, working with scientific collaborations, grassroots organizations, institutes and centers, community leaders, and funding agencies on the evolving HEP Software and Computing needs of experimental, observational, and theoretical aspects of the HEP programs. The scope should include research, development, maintenance, and user support.}
\end{addmargin}

\vspace{5mm}

Further details of the community vision for the CPSC can be found in the body of this report.

\end{tcolorbox}

Continued S\&C support for facilities, experiments, surveys, and theoretical calculations is essential for the health of the HEP science program.  This includes S\&C personnel as well as computing power, storage, and networking.  We have identified four key areas of need, listed below, where increased investment would significantly enhance the physics output of the US HEP community. \\

\vspace{-6mm}

Computing is a \textbf{global endeavor} and addressing the above items should include \textbf{coordination with worldwide partners}.  We also strongly support continued, significant investment in emerging technologies including quantum computing and machine learning, which were not part of the 2013 Snowmass process.  
\vspace{0mm}

\begin{tcolorbox}

\vspace{2mm}

\vspace{-4mm}
\begin{enumerate}
\setcounter{enumi}{0}
\item The US HEP community should take a leading role in {\color{blue}\textbf{long-term development, maintenance, and user support}} of essential software packages with targeted investment.
\begin{itemize}
\item A new structure is needed to fund modernization, maintenance, and user support of existing tools (grants typically only fund ground-breaking R\&D or development of new software).
\item Examples include (i) event generators and simulation tools like \texttt{Geant4}~\cite{GEANT4:2002zbu,Allison:2006ve,Allison:2016lfl} that do not belong to a particular facility, experiment, or survey, (ii) S\&C tools associated with one or more experiments, and (iii) data/software preservation after an experiment has ended.
\end{itemize}
\vspace{-2mm}
\end{enumerate}
\end{tcolorbox}

\begin{tcolorbox}
\vspace{-2mm}
\begin{enumerate}
\setcounter{enumi}{1}
\item Through existing, reshaped, and expanded %
programs, R\&D efforts {\color{blue}\textbf{cutting
across project or discipline boundaries}} should be supported from proof of concept to prototype to production. %
\begin{itemize}
\item Computational HEP is a vehicle for cross-cutting R\&D. Supporting research in this area at a variety of scales would be broadly impactful.
\item Examples include S\&C for theoretical calculations/generators; cosmological, accelerator, and detector modeling; machine learning methodology and hardware ecosystems; and algorithms and packages across experiment boundaries.
\end{itemize}
\vspace{-2mm}
\end{enumerate}
\end{tcolorbox}

\begin{tcolorbox}
\vspace{-2mm}
\begin{enumerate}
\setcounter{enumi}{2}
\item Support for computing professionals/researchers and physicists to conduct code re-engineering and adaptation will {\color{blue}\textbf{enable us to use heterogeneous resources}} most effectively.

\begin{itemize}
    \item Most HEP software runs on a single computing platform, making it difficult to use the multitude of hardware accelerators and diverse computing resources like cloud, HPC, etc.
    \item To satisfy the needs of inherently serial algorithms that are still transitioning towards computing accelerators or are not cost-effective to port, an appropriate level of traditional CPU-based hardware should coexist with more powerful heterogeneous resources.
\end{itemize}
\vspace{-2mm}
\end{enumerate}
\end{tcolorbox}

\begin{tcolorbox}
\vspace{-2mm}
\begin{enumerate}
\setcounter{enumi}{3}
\item Strong investment in {\color{blue}\textbf{career development}} for HEP S\&C researchers will ensure future success.

\begin{itemize}
\item Sustainable efforts in computation require continual recruitment and training of the HEP workforce.  We need to create an environment that is inclusive, supportive, and welcoming in order to integrate diverse skill sets and experiences.
\item Successful training events have been carried out through HEP experiments, institutes/organizations, and growing numbers of university courses.  We need to continue and grow these efforts for documentation and training at multiple levels.
\item Faculty/staff positions for physicists with expertise in S\&C for HEP are scarce and person-power shortfall in this area endemic. Funding agencies can catalyze faculty-level appointments in S\&C with joint appointments at national laboratories.

\end{itemize}
\vspace{-2mm}
\end{enumerate}

\end{tcolorbox}

\textit{Additional details for all of the above points can be found in the body of this report.}

\tableofcontents

\clearpage

\section*{1 Introduction}
\addcontentsline{toc}{section}{1 Introduction}

The Computational Frontier is different from many of the other frontiers that are part of the Particle Physics Community Planning Exercise (``Snowmass'').  For a number of the other frontiers, the goal is to identify large scale, long running projects that will be carefully considered by the Particle Physics Project Prioritization Panel (P5).  Our frontier, however, is important to almost all the other frontiers in that software and computing are ubiquitous in experiment and of essential importance in many areas of theoretical physics.  Two of the top ten most highly-cited papers of all time in particle physics\footnote{N.B. these are also only specific versions of these programs - the citation count is much higher if all versions are included.} (according to \href{https://inspirehep.net/literature?sort=mostcited&size=25&page=1&q=find%20topcite%2050%2B%20&ui-citation-summary=true}{\textsc{Inspire}}) are software programs~\cite{GEANT4:2002zbu,Sjostrand:2006za}. If we get the computing challenges right, the scientific output will follow.  If not, computing costs will be inflated or our productivity will suffer, and for some projects, it will suffer greatly.  For example, computing limitations (e.g., in terms of the ultimate precision) are on par with detector limitations in many flagship experiments --- without computing/methodological innovation, we will not be able to take full advantage of pristine data collected by state-of-the-art instruments.  

Another difference between computing infrastructure and a large experimental detector is that the latter often relies on custom or bespoke hardware, some of which may be expected to work for decades.  Computing hardware, on the other hand, is often replaced after half a decade (or less) and it is not yet clear what the next devices will look like.  However, the software in use in our community can last for decades, although it evolves over time.  In fact, over the last decade or so, changes in computer hardware have presented a major challenge to software development.  The issue of how to deal with heterogeneous nodes employing various hardware accelerator technologies does not have a single solution.  At this point, it is not clear which of several approaches will prevail, or even if there will emerge a single dominant approach.  This complicates the already challenging task of training for modern software development.  Coping with a rapidly evolving hardware environment is one of many challenges we investigate in this report. 


Despite differences between computing and instrumentation, there are also close connections.  Extracting physics from detectors requires experimental algorithms, simulations, and other software and computing tools.  For detector simulations and reconstruction algorithms, these tools are often tailored to specific instruments.  Even though traditionally, hardware is designed independent of software (which usually comes later), there is a growing movement towards codesign at all levels in order to maximize the physics output and minimize costs.  
For example, certain detector design choices may lead to computationally expensive simulation and reconstruction algorithms that may reduce the physics output volume and quality in practice and require increased investments in computing resources.  Furthermore, the physics sensitivity may even be enhanced with automated software algorithms, such as those based on machine learning.  Codesign is relevant for all areas of HEP, including theory (such as computing hardware and lattice gauge theory) and quantum information science (for HEP-tailored quantum computers).



\begin{figure}
    \centering
    \includegraphics[width=0.95\textwidth]{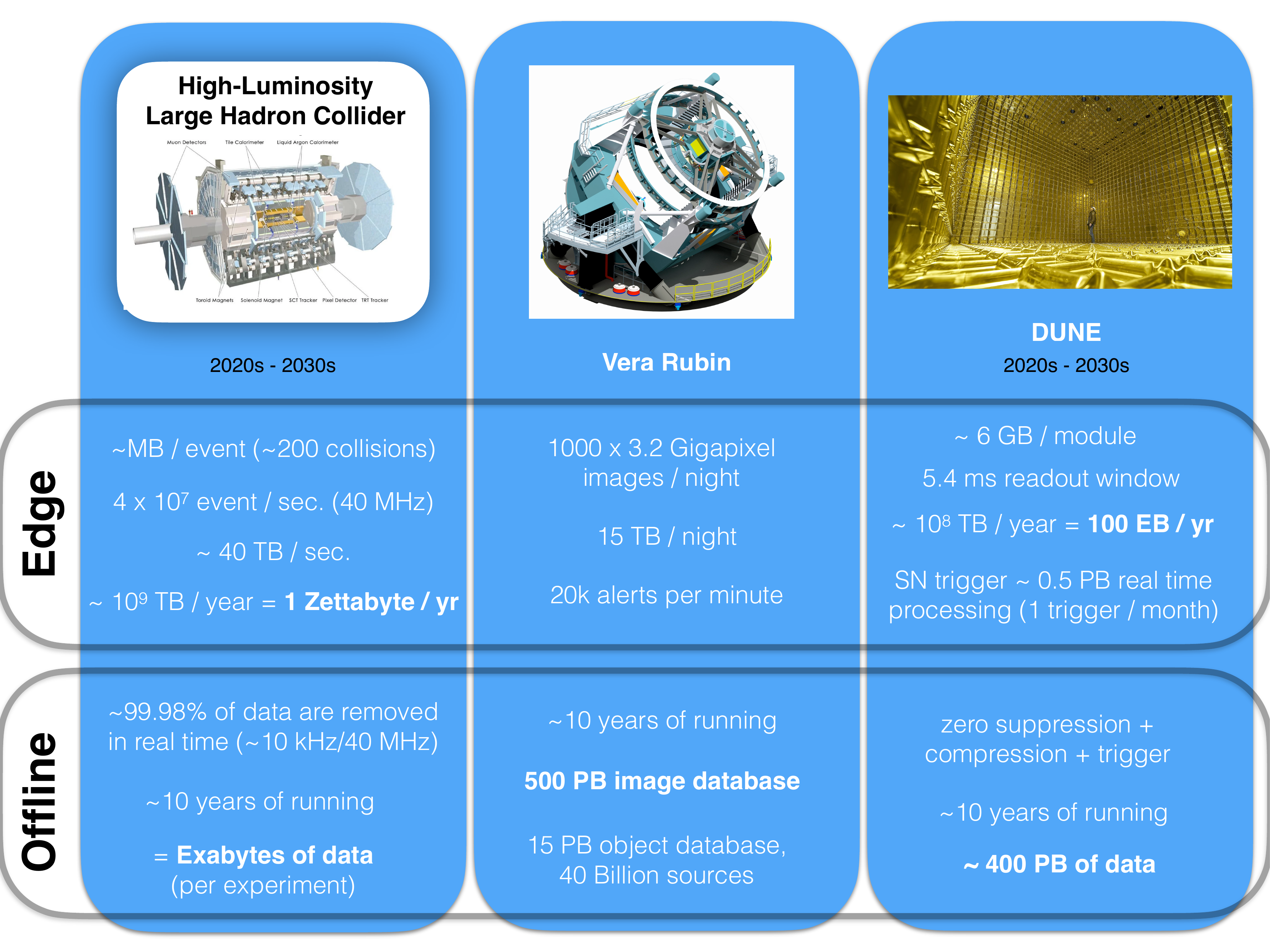}
    \caption{An overview of the online (low-latency or sometimes, `edge') and offline (high-latency) computing requirements for large experiments/surveys in the energy frontier ATLAS/CMS), cosmic frontier (Vera Rubin), and neutrino frontier (DUNE) over the next decade.  These are not all experiments/surveys and the numbers are meant only to be illustrative.  Furthermore, the challenges faced by many smaller experiments are not simply captured by such numbers.  See the rest of the report for further details.}
    \label{fig:scale}
\end{figure}

Our goal is to examine the software and computing needs of the particle physics community over the next 10 years, with a 20 year global vision for the field.  Within this timeframe, there will be a spectrum of experiments, surveys, accelerator R\&D, and theoretical efforts that will stress the community computational resources.  Figure~\ref{fig:scale} provides a brief characterization of some aspects of this challenge for big experiments across frontiers.  For example, the High-Luminosity Large Hadron Collider (HL-LHC) will run through the 2030s and its experiments produce an order of magnitude more data than during the initial LHC period.  Online (low-latency or sometimes, `edge') computing will become more prevalent and require various hardware accelerator systems to deliver low-latency decisions.  The cosmic and neutrino frontiers are not far behind the energy frontier in terms of data size.  Not captured by these numbers is the immense computational resources required for generating simulated data that are critical for theoretical predictions and data analysis.   We have mostly focused on the upcoming decade, where we need to address a resource gap of near-future programs and to account for the rapidly evolving technologies that make predictions beyond 2035 inaccurate. However, we also appreciate and consider the near-term R\&D program required to prepare for future facilities.

Software and computing have long been central to the HEP science program and therefore played a key role in the the 2013 Snowmass report~\cite{Bauerdick:2014qka}, but there are many new features of the frontier in this round.  The 2013 report was mostly organized along physics applications, with subgroups dedicated to computing for the Cosmic Frontier~\cite{Connolly:2013ibs}, Energy Frontier~\cite{Fisk:2014lia}, Intensity Frontier~\cite{Rebel:2013xaa}, Accelerator Science~\cite{Spentzouris:2013jla}, lattice field theory~\cite{Blum:2013mhx}, and perturbative field theory~\cite{Hoche:2013zja}.  Additional subgroups covered cross-cutting topics related to distributed computing and facilities infrastructure~\cite{Bloom:2013yva}, networking~\cite{Bell:2013fwa}, software development, personnel, and training~\cite{Brown:2013hwa}, and storage and data management~\cite{Butler:2013kka}.  For this round, we have forgone any organization by frontier and instead, we have set up topical groups that are only cross-cutting in nature.  This reflects the growing need for shared solutions to computing challenges that are not dominated by any one frontier.  To this end, the 2021--2022 Snowmass Computational Frontier (CompF) is organized into seven topical groups.  

The first topical group (CompF1) examined \textbf{Experimental Algorithm Parallelization}~\cite{compf1}.  Functional areas of this topical group include parallelization of detector reconstruction algorithms, physics object reconstruction/calibration algorithms, hardware utilization (including accelerators), developing better algorithms in addition to parallelization, and portability solutions that support the same algorithm implementation on multiple hardware architectures.  This includes software frameworks used for such tasks. 

The second topical group (CompF2) covered \textbf{Theoretical Calculations and Simulation}~\cite{compf2}.  This includes standalone perturbative and lattice calculations as well as theory calculations that are embedded in event generators.  Additionally, CompF2 considered detector simulations, particle accelerator simulations, and cosmic frontier simulations.  A cross-cutting issue was the utilization of hardware and the adaptation of software to hardware accelerators to take advantage of heterogeneous computing resources.

Topical group three (CompF3) explored \textbf{Machine Learning}~\cite{compf3}.  While multivariate analysis has been commonplace in particle physics for decades, modern machine learning has revolutionized everyday life and is poised to make paradigm-shifting contributions to particle physics.  Machine learning was essentially not a part of the 2013 report and so it is an emerging technology that has rapidly become a mainstream component of software and computing in particle physics.  CompF3 was tasked with understanding the development of methods and the deployment of machine learning methods at both small and large scales as well as in online and in offline settings.

A fourth topical group (CompF4) studied \textbf{Storage and Processing Resource} Access~\cite{compf4}, also known as Facility and Infrastructure Research and Development (R\&D).  Functional areas of CompF4 include access to data for large scale central workflows as well as end user analysis, access to high- and low-latency storage, access to CPU resources and various hardware accelerators, including specialized machine learning hardware, and the interconnection of these topics through networking.  While HEP has a long history of maintaining and utilizing centralized and distributed computing facilities, there is now a growing diversity of options.  High Performance Computing Centers (HPCCs) are becoming more heterogeneous, enabling and complicating experimental and theoretical workflows.  At the same time, the commercial cloud has significantly grown in capacity and accessibility which may open up the possibility of supplementing a non-trivial fraction of HEP computing.

The fifth topical group (CompF5) investigated \textbf{End User Analysis}~\cite{compf5}. This includes the development and utilization of analysis facilities as well as software topics related to analysis libraries (including \textsc{Root}~\cite{Brun:1997pa}), data storage formats, dataset bookkeeping, programming languages, software collaboration tools, and low- and high-latency analysis.  Since the last Snowmass, there has been a large transition from \texttt{C++}-based analysis to \texttt{python}-based analysis (see e.g., Ref.~\cite{Pivarski:2022ycs}).  
Additionally, the tool ecosystem has diversified due to the prevalence of excellent (non-HEP) community packages like \texttt{SciPy}~\cite{2020SciPy-NMeth}.  The HEP community is at a crossroads where we must decide to what extent we will maintain custom software for generic analysis or how much we will turn to or integrate (and contribute to) broader community standards into our workflows.

Topical group six (CompF6) covered \textbf{Quantum Computing}~\cite{compf6}.  Like machine learning, quantum computing was not part of the 2013 Snowmass process.  Due to rapid advances in algorithms and hardware, quantum computing has become a rapidly growing research area within high energy physics.  While fully error correcting quantum computers do not yet exist, noisy near-term devices still provide an exciting testbed for algorithm development.  The goal of CompF6 was to investigate the potential impact of quantum computing on particle physics and to determine what resource investment is required in the near term to explore the mid- and longer-term potential of quantum computing for HEP.  There is an intimate connection between quantum computing and other areas of Quantum Information Science (QIS), which are covered in other frontiers (TF10 in Theory and IF1 in Instrumentation).

Lastly, a seventh topical group (CompF7) explored \textbf{ Reinterpretation and Long-term Preservation of Data and Code}~\cite{compf7}.  This includes how/where various data products are accessed/stored.  Data products vary from the highly synthesized content of papers (including digitized figures) to the relatively raw public data that can be used for generic data analysis.  Central to this topical group are the tools for generating annotated public data and software, for combining results across experiments/surveys, and for archiving/re-running analyses (long) after the original research.

Not all topics are covered by our topical groups.  For example, training and career development as well as diversity and climate are cross-cutting challenges that are relevant for all topical groups.  There are also a number of topics that are shared between topical groups such as machine learning hardware at high performance computing centers (CompF3 and CompF4).  The topical group reports from this Snowmass process cover such topics from different viewpoints in order to explore cross-cutting challenges from the various stakeholders.

Before proceeding to the content of our report, it is useful to put the work of our frontier in the a global context.  Preceding this Snowmass process by a couple of years was the latest European Strategy Report~\cite{EuropeanStrategyforParticlePhysicsPreparatoryGroup:2019qin,CERN-ESU-015}.  Software and computing are prominently featured with recommendations such as: 

\begin{quote}
\textit{Both exploratory research and theoretical research with direct impact on experiments should be supported, including recognition for the activity of providing and developing computational tools.}

(...)

\textit{Large-scale data-intensive software and computing infrastructures are an essential ingredient to particle physics research programmes. The community faces major challenges in this area, notably with a view to the HL-LHC. As a result, the software and computing models used in particle physics research must evolve to meet the future needs of the field. The community must vigorously pursue common, coordinated R\&D efforts in collaboration with other fields of science and industry, to develop software and computing infrastructures that exploit recent advances in information technology and data science. Further development of internal policies on open data and data preservation should be encouraged, and an adequate level of resources invested in their implementation.}

(...)

\textit{Particle physics, with its fundamental questions and technological innovations, attracts bright young minds. Their education and training are crucial for the needs
of the field and of society at large. For early-career researchers to thrive, the particle physics community should place strong emphasis on their supervision and training. Additional measures should be taken in large collaborations to increase the recognition of individuals developing and maintaining experiments, computing and software.}
\end{quote}

Unsurprisingly, the recommendations of our frontier resonate with the European Strategy, with more details provided in the Executive Summary and Sec. 11.

\clearpage

\section*{\raisebox{-2.5ex}{\includegraphics[width=0.1\textwidth]{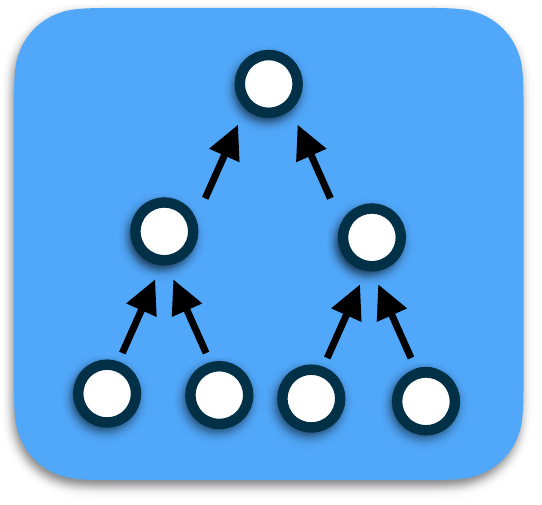}} 2 Experimental Algorithm Parallelization}
\addcontentsline{toc}{section}{2 Experimental Algorithm Parallelization}

Within HEP, experimental algorithms refers to elements of software developed to, for example, reconstruct physics objects (e.g., electrons, photons, muons, neutrinos, jets of hadrons, etc.), process observational data, and perform their calibration. Future experiments and cosmological observations will see a substantial increase in the volume of data collected, as compared to current programs. Additionally, the improvement of detector capabilities will have a significant impact on experimental algorithms, demanding more computing power and data storage in a world of flat budgets, where growth in resources will depend on the expected reductions in the cost of hardware. Although the focus of HEP experiments is typically high-throughput computing (HTC), in many cases provided by grid and local computing clusters, hardware evolution is prompting the community to increasingly utilize hardware-accelerated systems, which dominate High Performance Computing (HPC) facilities. These facilities are large US government investments which are being made increasingly available to the HEP community, while computing resources specifically dedicated to HEP become scarcer. Cosmic surveys and the accelerator modeling projects offer a good example of effective use have used HPC resources, having utilized leadership computing facilities for many years to carry out a large fraction of their simulation work. 

In most HEP scientific programs, the challenges presented by the increase in data volumes, as well as the rapid evolution of the computing landscape, compound with the time constraints imposed by the slow development cycles of experimental software, which is particularly acute in large collaborations. Consequently, the opportunities presented by emerging technologies, including HPCs and hardware specifically designed for AI/ML applications, offer the chance to bridge the gap between computing requirements and available resources. Success will, however, require continuous and significant R\&D investments during the next 10 to 15 years. The R\&D  program will seek to explore performance portability techniques to support multiple computing architectures while minimizing platform-specific code with minimal impact in performance. This program will also integrate AI/ML methods to tackle the analysis of complex data-sets to improve or preserve physics performance while reducing computational costs, and build computational frameworks that allow for collaborative software development.

Technology evolution will continue to demand significant investment in software frameworks within experiments, as algorithms are adapted to modern computing paradigms. In addition to parallelism on multi- or many-core CPUs, frameworks need to support execution of algorithms on heterogeneous platforms, including CPUs, GPUs, and field programmable gate arrays (FPGAs). Frameworks designed for multiple experiments, as well as common reconstruction tools, could significantly save resources by reducing code duplication. Common tools can provide ready-to-use software solutions, as well as a platform for R\&D work spanning usage of AI/ML methods, algorithm parallelization, and deployment at HPCs. This strategy would reduce the workload of individual collaborations, and benefit small experiments. 

In the particular case of hadron collider experiments, a critical challenge is to improve, or at least maintain, the physics performance of reconstructed physics objects in the context of growing numbers of hard collisions recorded in an event (pileup), as well as increased detector complexity. Additionally, charged particle trajectory reconstruction (tracking) is particularly expensive computationally and will benefit from more parallelism and the use of accelerators in heterogeneous platforms~\cite{HEPSoftwareFoundation:2020daq}. AI/ML has shown its power for physics object identification and has the potential to play an even larger role in particle reconstruction at future experiments. A convergence of trigger and offline algorithms, traditionally cruder and faster in triggers, seems feasible for the first time in history given the opportunities offered by hardware accelerators, such as GPUs. 

The field of cosmology has undergone a major transformation over the last two decades due to the advent of large sky surveys. For optical surveys, the algorithms can be broadly classified as falling into two classes: algorithms for static science where the data is collected over a long span of time, and time domain science, where algorithms are needed to find events quickly, classify them and provide guidance for further follow-up actions. In both cases, AI/ML algorithms and heterogeneous computing can play an important role, given the necessary forethought and support~\cite{2019arXiv190505116B}. In CMB experiments, the primary computational challenge is reduction of the large raw detector data to intensity and polarization maps, and then simulation of the instrument and map-making strategy. This will require use of accelerators and exploitation of a variety of computing systems, including HPCs and distributed resources. Direct dark matter detection experiments face a different set of challenges with regard to experimental algorithms than the cosmological surveys. Current experiments are approaching data volumes of $\sim$1 PB/year, see, e.g., Ref.~\cite{Mount:2017qzi}. Their challenges more closely mirror those of collider physics experiments in what relates to the use of HPC resources and scalable HEP frameworks, as well as AI/ML techniques. 

Experimental algorithms for neutrino experiments, and, in particular, for liquid argon time projection chamber (LArTPC) detectors, can be divided in two big categories: signal processing and high-level algorithms. Signal processing algorithms operate on the waveform signal from the TPC. Higher level algorithms rely on the signal processing for physics object clustering and identification, including tracking, calorimetry, and particle identification tasks. Different methods are being employed by experiments or investigated in the context of more general R\&D projects, using techniques which span from traditional algorithms to deep learning. Computing challenges arise from the large amount of raw data for a full event readout, imposing stringent demands on memory usage and disk space~\cite{DUNE:2020ypp}. Parallel processing patterns must continue to be employed, using both CPU-only resources and mixed CPU-GPU. Software packages such as the Wire-Cell Toolkit~\cite{Qian:2018qbv}, designed for highly multi-threaded operations, must continue to be improved and supported. Continuous development of the common LArSoft library is vital to enable direct sharing of developments across currently operating LArTPC experiments~\cite{FASER:2022yqp}. R\&D to run data processing tasks for neutrino experiments at HPCs exploiting optimization features at the node level, such as vectorization and multi-threading, need to be undertaken.

Experiments targeting rare events are also implementing algorithms to run on heterogeneous architectures, including at the trigger level, achieving significant savings in computing power. This path has been followed by experiments exploring flavor physics as well as muon experiments.

All these challenges presented by experimental algorithms across physics frontiers call for a common set of recommendations: 

\begin{itemize}
    \item Workable solutions must be found to exploit {\bf heterogeneous computing platforms}. Portability solutions must be developed and supported. In parallel, targeted optimizations of key algorithms for experiments often provide the largest speedups and must be supported.
    \item {\bf Software frameworks and common tools} must evolve and adapt to the new computing landscape to enabling the usage of parallel algorithms in production environments of big and small experiments.
    \item {\bf Interdisciplinary collaborations and programs} should play a critical role in developing algorithm approaches to take full advantage of emerging hardware and software technology.
    \item {\bf Training opportunities} for early-career researchers are essential to ensure that the experiments acquire the necessary expertise to utilize these technologies.
    \item {\bf Career opportunities} targeting researchers who focus on experimental algorithm development need to be created. The scientific value of this work is typically not fully appreciated in faculty and staff searches.
    \item To enable long-term sustainability of HEP software products and offer job security to researchers in this area, {\bf software development should be supported} according to a model similar to that for detector projects, or directly made part of detector projects. Funding needs to go beyond the R\&D phase of algorithm and software exploration and into the development, production, and maintenance phases.
    \item Funding agencies and research teams have to commit {\bf sustainable long-term effort to mission-critical software products}.
    \item Big experiments/surveys need {\bf multi-year computing resource allocation planning} for use of national facilities, as an alternative to the current model of annual allocations. 
\end{itemize}
    
\section*{\raisebox{-2.5ex}{\includegraphics[width=0.1\textwidth]{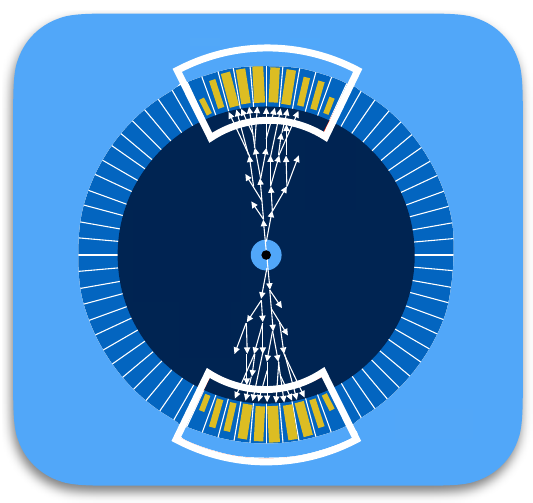}} 3 Theoretical Calculations and Simulation}
\addcontentsline{toc}{section}{3 Theoretical Calculations and Simulation}

Theoretical calculations and simulation overlap with almost all aspects of HEP and almost every Frontier. 
The topic is very broad and was divided into six domains for the purpose of this report: cosmic calculations, particle accelerator modeling, detector simulation, event generators, perturbative calculations, and lattice quantum chromodynamics (QCD).
Computational challenges are different across these domains. For
instance, accelerator simulations require a highly coordinated interplay of several tasks, which benefit from fast inter-node connections available in HPCs, while matrix element calculations for different phase space points in event generators are completely independent of one another and can be executed in separate nodes in a computer network, including HPC facilities. Nonetheless, we have identified significant commonality in computing needs among the six areas, which we list alongside those specific to a domain.

\begin{itemize}

\item {\bf Hardware accelerator-friendly, portable common tools} are essential to adapt common software packages to run efficiently on hardware accelerators, including GPUs, and should continue to be supported with sufficient effort to deliver complete products and incorporate portability solutions to support different coprocessor devices and architectures.
\item {\bf Universal programming interfaces} would remove barriers to portability and return on investment in software.
\item {\bf Early access to new computing hardware} is a key element to ensure timely scientific exploitation.
\item {\bf Right-sized CPU clusters} are necessary to provide general purpose computational cores with high performance memory for important algorithms that do not easily map to computational accelerators.
\item {\bf Automation of memory hierarchy} would remove the significant burden of explicitly moving data between different levels of the memory hierarchy, avoiding a real drain on scientific programmer productivity.
\item {\bf Best practices for common software} should be encouraged to ensure interoperability and common data structures. Existing standard libraries and underlying software technologies should be used when available and feasible.
\item New processes and organizations should be established with funding agencies for {\bf long-term software maintenance and support of common software tools} in the areas of accelerator modeling, detector simulation, and physics generators must be strengthened.
\item {\bf Collaboration with communities of computer scientists, applied mathematicians, nuclear physicists and others relevant to HEP computational and physics model development efforts} should continue to be supported and encouraged in the context of DOE and NSF programs designed to create common computing and simulation infrastructure, as well as to achieve effective utilization of U.S. resources across scientific fields. 
\item Programs of {\bf joint lab-university tenure track appointments in the area of S\&C} should be created with the goal to help universities build strong research groups focused on topics of S\&C for HEP, while generating academic opportunities for young researchers in that area.
\item {\bf Training} in both novel architectures and AI/ML with a low barrier to access for graduate students, postdoctoral researchers, and domain scientists is critical to ensuring the skills base exists for productive science in theoretical calculation and simulation.

\end{itemize}

\subsection*{3.1 Cosmic Calculations}

In order to successfully achieve the observational goals, a state-of-the-art simulation and modeling program is needed~\cite{Alvarez:2022}. Cosmological simulations provide an invaluable tool to optimize the observation design, to interpret observation data, and to help unearth the underlying physics. Next-generation cosmological simulations will increasingly focus on physically rich high-fidelity simulations that can directly connect with observational outputs from multiple different surveys. 

Several types of cosmological simulations are employed. The first-principles N-body simulations that have no free parameters can accurately describe dark matter fluctuations from the largest observable scales down to scales deep into the nonlinear regime. Additional models such as the halo occupation distribution,
sub-halo abundance matching, or other schemes are added on top of a simulation to reconstruct galaxies. Hydrodynamical simulations provide a reasonably accurate description of the distribution of
baryons, quantify the effects of baryons on various probes of large-scale structure, and provide useful results for the distribution and properties
of galaxies and clusters. 

The diverse observational probes present significant challenges in cosmological simulations.
One challenge is the computational cost to include multiple probes with 
high precision. In addition, computationally demanding hydrodynamical simulations are required to simulate some observables such as the thermal and kinetic Sunyaev-Zel'dovich effects, tSZ and kSZ, to infer the distribution of gas in the universe, or Lyman-$\alpha$, to constrain thermal properties of the intergalactic medium. Semi-numerical
simulations are used when the hydrodynamical simulations
cannot cover large enough volume while reaching small enough halos.
Another challenge for cosmological simulations is to 
reproduce observations, for example, correlations
between multi-wavelength observables, in consistent galaxy formation models. It is necessary to develop simulation models that can be applied to
gravity-only simulations and account for correlations between neutral and ionized gas, stars, and dust in galaxies and galaxy clusters. Also, it is not trivial to simulate massive neutrinos since they make up a non-negligible fraction of the total energy but can be decoupled
when relativistic and have a free streaming scale.
In the N-body simulations, thermal velocity distributions
need to be accounted for in the structure formation calculation on smaller scales. Covering both the large volume required by the weak lensing and maintaining the accuracy at small scales required by strong lensing presents another computational challenge in ray tracing. In the modeling for future analyses of small scale measurements, baryonic feedback effects that can change the local matter density must be included in the cosmological simulations. There is also the need to include models of modified gravity for dark energy studies, as well as
a variety of dark matter models (warm dark matter,
interacting dark matter, self-interacting dark matter, ultralight dark matter, ultraheavy dark matter, multiple component dark matter~\cite{Banerjee:2022}). Additionally, neutrino experiments demand modeling of supernovae. Next-generation cosmological surveys demand better understanding, mitigation, and control of systematic uncertainties using realistic ``virtual universes'' from cosmological simulations. 

Cosmological simulations are computationally intensive and demand the use of state-of-the-art computer hardware. With the arrival of exascale supercomputers, cosmological simulations will need to use computer accelerators and memory access patterns effectively, as well as performance portable programming models and scalable algorithms. Scalable analysis approaches are of utmost importance, since the handling and processing
of large simulation data sets require large computing resources in their own right. Furthermore, co-development of simulation software and analysis tools is recommended for a successful cosmological simulation program.  

\subsection*{3.2 Particle Accelerator Modeling}

Particle accelerators are large and complicated scientific devices that can be thousands of meters long with hundreds of thousands of elements. Their design, construction and operation are both sophisticated and expensive, and depend critically on computer modeling. To meet the needs of next-generation accelerators, it is important for modeling tools to
include all types of accelerating structures (e.g., RF-based,
plasma-based, structured-based wakefield, plasmonic) and different accelerator components (e.g., superconducting magnets, structured plasmas), and to target the accelerator and beam physics thrust grand challenges on intensity, quality, control, safety, and prediction. 

The final particle beam quality delivered by particle accelerators depends on the performance of the entire accelerator system, consisting of many different segments.  To attain an optimal design, end-to-end simulations starting with the source and ending at the target or the interaction region are needed. These simulations should include first-principles models for high precision accelerator design and fast machine learning-based surrogate models for online particle accelerator tuning. Such a virtual particle accelerator can be used as an emulator of the real physical accelerator in the accelerator operation. One challenge in the end-to-end simulation is to transfer particle data from different components of the particle accelerator~\cite{openPMDstandard}. High-fidelity end-to-end accelerator simulation is computationally very expensive, especially when including collective effects, with run times varying from days to months. Cutting-edge and emerging computing techniques including advanced algorithms, AI/ML methods, and quantum computing may substantially 
reduce the computational time and enable a fast real-time optimization with end-to-end simulations. 
Advanced algorithms play an important role in accelerator modeling.
High-fidelity self-consistent simulations cannot be achieved without advanced algorithms, which can lead to significant improvements in computational speed~\cite{Qiang:2006,Qiang:2009}. R\&D is needed to explore new algorithms that exhibit better properties, remove the bottlenecks in existing algorithms on cutting-edge heterogeneous computing hardware (e.g., GPU, FPGA), and improve their speed and accuracy. Differentiable simulations show great potential to reduce the number of simulations needed for optimization and design. Using AI/ML surrogate models of the physics simulation can save orders of magnitude computing time~\cite{Edelen:2020}. Support is needed for the development of AI/ML modeling techniques and their integration into accelerator simulation and control systems, with an emphasis on continual and adaptive learning for time-varying systems, uncertainty quantification, and physics-informed methods to enable broader model
generalization to new application conditions.

Particle accelerator modeling is computationally intensive and has utilized supercomputers for many years. It has used both high-performance computing to model collective effects and high-throughput computing to model single particle dynamics. On exascale supercomputers and special purpose computers with hardware accelerators (e.g., GPUs), it is a challenge to achieve the supercomputer peak performance due to deep levels of independent memory caches, on top of global and neighboring communications that are used to exchange information across multiple computer nodes during the simulation. As part of this, the workload needs to be uniformly distributed among many compute units to achieve good parallel efficiency. Support is needed to adopt portable, state-of-the-art programming paradigms and frameworks that support both multi-level parallelization and effective dynamic load balancing over all levels of compute parallelism, so that simulations run efficiently on computer architectures with scalable I/O, post-processing and in situ data analysis solutions. Furthermore, it is desirable to organize the beam and accelerator modeling community in open efforts to foster the development of ecosystems of interoperable codes and tools, libraries and frameworks, based on API and data standards, establish open access data and workflow repositories, and have an organized governance structure. Finally, support for development, maintenance, data-driven validation, user support and training are essential to guarantee sustainability, innovative potential and reliability of particle accelerator simulation.

\subsection*{3.3 Detector Simulation}

Accurate detector simulation is crucial for tasks ranging from detector design to data analysis~\cite{Frontiers:futSimu2022}.
\texttt{Geant4}~\cite{GEANT4:2002zbu,Allison:2006ve,Allison:2016lfl}
is the primary tool used for detector simulation throughout HEP~\cite{Banerjee:2022jgv}.
The \texttt{Geant4} collaboration and the HEP community engage in continuous research and development to improve the computational performance, refine the models of physical interactions, and incorporate new technical features and models.
\texttt{Geant4} is the center of an ecosystem that includes numerous software packages serving different needs. This includes material modeling and geometry navigation, the physics of the interaction of particles with matter, as well as fast simulation engines and prototypes to execute tasks on GPUs.

The next generation of collider experiments, including the HL-LHC and the various future colliders proposed during the Snowmass process, will pose extreme computing challenges for detector simulation.
Compared to present-day experiments, data volumes will grow by at least an order of magnitude. Additionally, detectors will incorporate novel technology to increase the precision of physics measurements, thus demanding more detail in geometry descriptions and accuracy in physics models. The number of events in \texttt{Geant4}-based simulation samples will therefore have to increase accordingly, and so will the fidelity of its physics modeling. During the LHC Run 2 for CMS and Atlas, \texttt{Geant4} used 40\% of all computing resources~\cite{HEPSoftwareFoundation:2018fmg}, while for LHCb the corresponding number is ${\sim}70\%$~\cite{Muller:2018vny,Bozzi:2802074}.
For CMS, it is projected that only 14--20\% will be available for simulation at the HL-LHC~\cite{CMSOfflineComputingResults,Collaboration:2802918}.

In addition to high energy colliders, experiments of various sizes that search for light or weakly interacting particles rely on \texttt{Geant4}~\cite{FASER:2022yqp,Kahn:2022kae,Roberts:2022ezy}.
A wide range of other  experiments face similar computational challenges to process petabytes of data, including the corresponding simulation of background and signal processes across a wide range of energy scales, from TeV muons to optical photons traversing meters of complex materials. 
The need to meet these challenges with substantially less personnel makes the development of common software even more important. Future neutrino, dark matter, and other similar experiments will obtain even larger data volumes and probe even more precise physics, increasing the computational challenges and the associated need for high-performing and well-supported software.

The increasing breadth and precision of next-generation HEP experiments will require computationally expensive improvements in the physics models used in \texttt{Geant4}~\cite{Banerjee:2022jgv}. The future of detector simulation software is therefore turning toward the use of GPUs, targeting, in particular, HPC facilities that are intended to provide a larger fraction of HEP computing for next-generation experiments~\cite{Frontiers:futSimu2022}. Because detector simulation is naturally an HTC problem, nontrivial effort is required to adapt the necessary computations to use computer accelerators effectively. Two ongoing projects, AdePT~\cite{AdepT} and Celeritas~\cite{johnson_2021,Celeritas:SM}, the latter led by US national laboratory personnel, have produced viable prototypes of high-performing GPU-native simulation engines. However, there is a significant burden in developing and maintaining coprocessor-friendly code, including portability to different GPUs platforms and, potentially, future non-GPU architectures. Another avenue to achieve faster simulation is the use of machine learning to replace or augment existing ``classical'' (rule-based) simulation engines~\cite{Adelmann:2022ozp}.
Full replacement is commonly pursued using generative models such as generative adversarial networks (GANs), variational autoencoders (VAEs), or normalizing flows (NFs), although achieving an acceptable level of physics fidelity presents a significant challenge. Augmentation to improve lower-quality results, sometimes called refinement, can be applied to classical fast simulation engines or to generative models, using techniques similar to those above or regression-based approaches.
The possibility also exists to use coprocessors more cost-effectively and efficiently by performing the ML inference as a service, avoiding the need for each CPU to be equipped with a GPU~\cite{Harris:2022qtm}.

The above solutions, including GPU prototype transport engines and ML algorithms, are promising avenues
for faster execution of detector simulation. However, it is unlikely that these will completely supplant \texttt{Geant4}, which provides broad support for a diverse set of use cases.
Therefore, some percentage of general purpose von Neumann CPU-based computing will be needed in HEP for the decades to come.
There is a strong consensus throughout the field that the decline in  detector simulation investment, including the \texttt{Geant4} toolkit, must be reversed, with funding not just restored, but increased above historical levels.

\subsection*{3.4 Physics generators}

The physics generator software landscape is complex and varied because it needs to provide solutions to different aspects of event generation:
event (matrix element) calculation, hadronization and parton shower modeling, underlying event tuning, parton matching and merging, particle decays, cross section calculations, parton distribution functions, data formats, and analysis and reinterpretation.
There are many software packages that address some of these needs, which often include plugins to perform specific calculations and provide ways for users to plug in their own custom models and computations.
Physics generation is more or less computationally expensive, depending on the precision of the calculation. Overall, the percentage of grid computing time used for event generation at the CMS and ATLAS experiments is estimated to be 5--12\%, with similar or even more substantial requirements expected for ALICE and LHCb~\cite{HSFPhysicsEventGeneratorWG:2020gxw}.
Moving from LO to NLO and higher orders, or increasing the multiplicity with additional partons~\cite{Hoche:2019flt}, typically causes a factorial increase in CPU usage for the same processes, while different processes may still have very different baseline computational requirements.
The memory usage and multithreading abilities of different generator software are also highly variable and can lead to inefficiencies.
Generator CPU usage is projected to increase to 8--20\% in the HL-LHC era, when higher precision will frequently be required~\cite{CMSOfflineComputingResults,Collaboration:2802918}.
The generator usage at neutrino frontier experiments has not been measured and profiled as systematically, although we know that the computing budget is generally smaller compared to the energy frontier.
However, the increasing precision of upcoming neutrino experiments such as LBNF/DUNE and HyperK is expected to require similar increases in experiment computing time devoted to event generation.
Further future experiments will have different generator needs depending on their initial states.

Physics generators software faces challenges, especially when it comes to keeping up with the increasing demands for both physics precision and computing efficiency.
The automated computation of NLO and higher order diagrams, introduces events with negative weights, which can substantially reduce the statistical power of the final result.
There are also inefficiencies in phase space sampling, e.g., in slicing, biasing, or filtering, to change kinematic features, and in merging to avoid double-counting between matrix element generators and parton shower modelers.
Further, use of generators by experiments may involve significant repetition of expensive calculations, especially to assess certain systematic uncertainties that cannot be represented as alternative event weights.
Historically, there has not been a single obvious locus to coordinate common activities, even if MCNet (\url{https://www.montecarlonet.org}) and, more recently, HSF have provided useful forums, especially in the areas of training and computational efficiency, respectively. In the future, more emphasis should be placed on establishing and coordinating common activities.
Projects to reduce the computational burden of event generation, for example by adapting to use GPUs, need a substantial increase in effort in order to be successful. An opportunity comes from the possibility to parallelize the matrix element calculations, which take up more than most of the computing time for complex LHC physics processes. Matrix element calculations are a perfect fit for CPU vectorization and hardware acceleration on GPUs. 
The use of ML for approximate matrix element calculations, including phase space integration and sampling as well as fully generative models is also very promising, but still in a relatively early stage of R\&D. Additionally, there are efforts currently underway to explore the use of ML for parton shower modeling, hadronization, event (un)weighting, and tuning~\cite{Campbell:2022qmc}.
Some of the difficulties that generators face in computing are intrinsically tied to the nature of the calculations being performed, so fundamental research may be required to develop better methods.
Overall, event generators are usually supported by small teams that frequently have little to no dedicated funding and cannot adequately plan for succession. As with detector simulation, it is vital to the entire field that funding is provided for permanent positions to support, improve, and expand these crucial components of HEP software. It has been suggested that a diverse and cross-cutting collaboration, similar to MCNet in Europe, could bring together the U.S. event generator community to share resources and ideas~\cite{Campbell:2022qmc}.

\subsection*{3.5 Continuum Field Theory Calculations}\label{sec:perturb}

Continuum field theory calculations include both precision high order perturbative calculations and conformal bootstrap theoretical calculations. Precision is important to future experiments, where electroweak processes are moving from percent scale accuracy down to per-mille accuracy. This sets the goal for precision electroweak phenomenology on a range of processes. At the high luminosity LHC, even rare processes like Higgs production require theoretical control of cross sections at the 1\% level~\cite{Cordero:2022gsh}. 
Since the 2013 Snowmass exercise, the state-of-the-art has advanced dramatically and now next-to-next-to-next-to-leading order N$^3$LO QCD calculations for $2\to 1$ processes and NNLO QCD calculation for $2\to 3$ processes and $2\to 2$ processes with internal masses are becoming state-of-the-art and reasonably fast on  contemporary computing resources, while automation and more complex final states at NNLO and diboson production at N$^3$LO are goals for the Snowmass period~\cite{LesHouches2021}. 

The computational challenge is to perform Feynman loop integrals within perturbative quantum field theory, giving access to the scattering matrix and its analytic structures, often involving non-standard special mathematical functions. Numerical integration software is increasingly able to use GPUs. New semi-analytic methods for solving master integrals have been developed and reduce evaluation times to order of a minute per phase space point for two loop integrals. QCD corrections to scattering amplitudes have been fully automated at NLO using many public and private tools. A core computational bottleneck is in the handling of linear relations between Feynman integrals (integration by parts identities) to reduce many graphs to a limited set of master integrals. From a computational perspective the use of modular arithmetic in the calculation of integration by parts reduction is an interesting transformation of the computational requirement that may require radically different computing hardware solutions. These calculations require a significant amount of memory per core. The development of codes for challenging amplitude calculations in a HPC friendly way may have large memory demands and require run-times that are possibly well beyond available batch limits on a given cluster. Projected computational costs of different multi-loop calculations envisioned in the Snowmass period extend to as much as 10M CPU core hours in some cases~\cite{Cordero:2022gsh}.
Although there is some role for GPUs in numerical integration, they only match a modest part of the scientific requirement~\cite{Cordero:2022gsh}. Large memory and runtime limitations imposed by typical HPC environments mean that bespoke computing nodes and queues are required to enable these calculations. Machine learning does not at this time
play a large role, although some applications have been used in event generation.

The idea of the conformal bootstrap is to constrain and solve conformal field theories (CFTs) using physical consistency conditions like symmetry, unitarity, and causality. By relying on nonperturbative structures, bootstrap methods can work even in strongly-coupled systems where traditional perturbative techniques fail. 
From a computational perspective, the bootstrap methods suffer two key algorithmic challenges: the need to  find faster optimization methods and efficient methods for exploring high-dimensional spaces of CFT data.
Distributed high precision arithmetic which can be parallelized over many HPC nodes and hundreds of thousands of cores is required, leveraging new hardware like GPUs and FPGAs.

Both multi-loop perturbative calculations and numerical conformal bootstrap
have a similar computer hardware requirement. They both have a current
software reliance on CPU cores, and a large memory per core requirement.
Perturbative calculations have a sizable barrier to the use of GPUs,
while the conformal bootstrap approach, after significant software engineering, may be amenable to GPU or FPGA acceleration of high precision arithmetic.

\subsection*{3.6 Lattice QCD}\label{sec:lattice}

Lattice gauge theory provides a systematically improvable numerical evaluation of the Euclidean Feynman-path integral for making first principle predictions of Quantum Field Theories. Worldwide lattice gauge theory efforts directly support numerous high-energy physics experiments by calculating properties of hadrons that are vital to interpretation of many experiments. These efforts make significant use of supercomputers and are critically dependent upon continued computing advances. Today, simulations of Lattice QCD provide precise model-free theoretical predictions of the hadronic processes underlying many key experimental measurements: for example determining the hadronic contributions to the anomalous magnetic moment of the muon (Muon g$-2$)~\cite{Muong-2:2021ojo}, understanding of nucleon structure and parton physics, contribution of CP violating operators (in SM and BSM) to neutron electric dipole moment, the form factors for neutrino-nucleus interaction (critical to DUNE), quark-flavor physics, anomalies in the B sector (LHC, Belle II), in reconciling the CP violation observed in kaon experiments with the SM, and exploring rare kaon decays~\cite{https://doi.org/10.48550/arxiv.2204.07944,https://doi.org/10.48550/arxiv.2202.07193,https://doi.org/10.48550/arxiv.2203.10998}. A critical review of important lattice QCD predictions that are considered mature with respect to control over all sources of errors is carried out by the Flavour Lattice Averaging Group~\cite{Aoki:2021kgd} every two to three years to inform the larger community of progress and predictions.

Achieving the precision in lattice QCD calculations to satisfy the needs of the experiments requires simulations at finer lattice spacing and larger volumes, which needs further algorithmic research, novel computer hardware design beyond the exascale, improved software engineering, and attention to maintaining human resources~\cite{Boyle:2022ncb}. Present algorithms display growing limitations as substantially greater ranges of energy scales are included in the  problem, an algorithmic challenge called critical slowing down. The development of numerical algorithms is a significant intellectual activity that spans physics, mathematics, and computer science under investigation in the context of exascale computing, ML including tensor networks, and quantum computing.

The massive vector parallelism of lattice gauge theory is, 
in principle, amenable to GPU and possibly other acceleration.
This imposes a significant additional programmer overhead and  the need for the most commonly used packages to receive sufficient investment (GRID~\cite{Boyle:2022nef,Boyle:2016lbp,Boyle:2017gzg}, MILC~\cite{DeTar:2018pyj,Gottlieb:2000tn}, CPS~\cite{Jung:2014ata}, Chroma~\cite{Edwards:2004sx} and QUDA~\cite{Clark:2009wm}). For smaller projects, especially where rapid development and programmer productivity are at a premium, it is better and more cost effective in terms of human effort to maintain access to a range of CPU resources. USQCD institutional clusters at Brookhaven National Laboratory, Fermilab, and Jefferson Laboratory have been instrumental in supporting the significant number of smaller and experimental projects that would not
achieve the return on investment to justify bespoke software development for multiple architectures.

\section*{\raisebox{-2.5ex}{\includegraphics[width=0.1\textwidth]{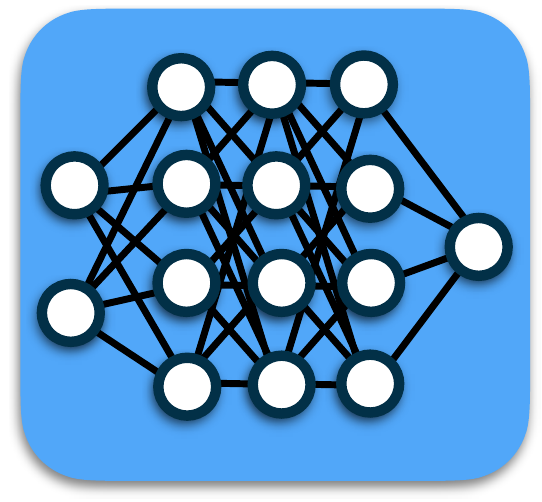}} 4 Machine Learning}
\addcontentsline{toc}{section}{4 Machine Learning}

Multivariate analysis has a long history in HEP.  Many of the discoveries made in the last decades were enabled, accelerated, or otherwise improved through the use of machine learning (ML) techniques.  For example, ATLAS and CMS have published over 600 papers searching for new particles and the \texttt{TMVA} multivariate analysis package~\cite{Hocker:2007ht} is cited in over 10\% of them.  This does not include the analyses that use per-object ML-based classifiers (such as flavor tagging), which are likely a large fraction of all searches.  Modern machine learning and deep learning in particular have the potential to significantly improve performance across the board.  At the same time, these novel approaches have created opportunities to optimize research in qualitatively new ways.  Additionally, cross-cutting developments have built bridges within the HEP community as well as with other areas of science and with industry.  These facts have justified the inclusion of a dedicated ML topical group in this Snowmass process for the first time.

We have identified a number of areas that require dedicated resources in order for ML in HEP to flourish over the next years:

\paragraph{Dedicated HEP-ML Research} While industry is driving many of the developments in machine learning, HEP has unique challenges that require dedicated solutions.  For example, the level of precision we require and the use of simulation-data hybrid methods for inference call for HEP-specific research in the areas of uncertainty quantification (UQ), validation, and interpretability.  Furthermore, HEP data have a complex structure that can benefit from physics-aware learning with custom ML architectures, learning strategies, and/or evaluation metrics.  This structure includes non-trivial geometry/topology/enumeration (vector, set, list, image, tree, graph, point cloud, etc.) as well as invariance/covariance under various symmetry groups.  One way to engage the broader ML community in these research topics is with the development of well-curated public benchmark datasets. 

There are also a number of HEP research areas where machine learning has traditionally not played a role, but where modern machine learning has the potential to be transformative.  One such area is `particle theory'\footnote{Quotes used here because ML method research blurs the lines between `theory' and `experiment' in a productive way that draws on insights from both communities.}, including simulation and lattice gauge theory.  A class of ML methods called generative models are able to emulate complex probability densities in high dimensions, which can be used to augment or accelerate various steps of the HEP simulation chain and also may solve key challenges in lattice gauge theory.  The research area called \textit{simulation-based inference} (also sometimes synonymous with \textit{likelihood-free inference}) is the optimal combination of simulations with ML and may enable analyses that are not possible with classical methods.  In particular, such tools allow for exploring the data and model parameter spaces in high dimensions moving from a multivariate approach to \textit{hypervariate} analysis. 

The opposite extreme is \textit{anomaly detection} and other \textit{label-free} learning strategies where ML may be used to reduce the reliance on specific HEP models.  Such tools provide an orthogonal approach to traditional analyses and may be essential for broadening the HEP search program for new physics.  Most HEP analyses are constructed with narrow hypothesis and anomaly detection may be a complementary tool for exploratory analysis.  More generally, there is a spectrum of model dependence and ML methods can be used across this continuum and for both achieving signal sensitivity as well as estimating Standard Model backgrounds.

Another significant growth area for ML and HEP is instrumentation.  ML methods can be used for all aspects of detector research, including experimental design, quality assurance/control, and calibration.  A key insight from dedicated HEP-ML research is to cast HEP problems as ML problems and then use the most effective optimization strategies available.  For example, gradient descent is one of the most efficient optimization procedures and experimental design can be optimized in this way if detector simulations are made differentiable (through \textit{differentiable programming}) or if a differentiable model (called a \textit{surrogate model}) such as neural network is trained to emulate a detector.  There is a direct connection between how an ML problem is set up and the physics outcome of the optimization.  Using out-of-the-box approaches may lead to undesirable properties such as prior dependence, bias, etc.

Additionally, ML may be useful for control at experimental facilities such as particle accelerators.  Ultra low-latency inference is relevant for facilities and experiments and this is another area  where HEP can contribute approaches that may be useful more generally.

\textit{Many of the topics mentioned above are inherently interdisciplinary and so funding structures that are inherently cross-cutting are essential for success.}  Recent interdisciplinary institutes from DOE and NSF that involve HEP are excellent examples of cross-cutting structures.

\paragraph{Software and Hardware Needs} For many years, \texttt{TMVA} was the most common ML tool in HEP.  Now, there is a more diverse ecosystem with many people relying on industry standards such as \texttt{SciPy}~\cite{2020SciPy-NMeth}, \texttt{TensorFlow}~\cite{tensorflow2015-whitepaper}, and \texttt{PyTorch}~\cite{NEURIPS2019_9015}. All of these tools are open source and we should support HEP software specialists to directly contribute to these projects to ensure that HEP needs are met and to give back to the broader community.  Incentives could be created for code sharing and reproducibility through expectations from funding sources.

In addition to a heterogeneous software ecosystem for ML, there is also a diverse set of hardware tools available for modern machine learning.  Graphical Processing Units (GPUs) are essential for enabling a reasonable development cycle.  While GPU resources are available from many computing centers, it is often invaluable to to have a dedicated GPU system for prototyping.  Research grants funding ML could be even more effective if they also carried a small hardware component. For larger tests and for deployment, high performance computing centers are essential.  This requires allocation systems to be amenable to R\&D and for computing systems to be compatible with standard software packages so that the typical HEP researcher does not need GPU programming experience.

There is significant R\&D in industry for developing even more advanced hardware accelerators for machine learning training and inference.  The HEP community should stay connected with developments in these areas through direct connection to industry and through computing centers.  Additionally, the HEP community may have more stringent latency and radiation tolerance requirements for certain inference tasks (trigger systems, operations in space, etc.) that may require custom solutions.  Development in this area will require collaboration with instrumentalists within and outside of the HEP community.  Lastly, training and deployment of machine learning algorithms may benefit from cloud and other on demand services, which may reduce the need for custom hardware to be on premises.  The exact portfolio of HEP-funded hardware investments versus using community resources should be optimized.

\paragraph{Training and Personnel} The ubiquity of ML for all areas of HEP science means that these tools should be part of the standard training program.  This could take many forms, including graduate school courses on ML as well as summer schools.  There are a growing number of examples for both of these topics and the community could benefit from sharing ideas/experiences in order to refine curricula for the future.  HEP could consider joining/supporting or otherwise creating training materials similar to what the APS Group on Data Science is organizing through its \href{https://dsecop.org}{Data Science Education Community of Practice (DSECOP)}.

In addition to broad training, we also need to have a training path for researchers who want to develop new methods at the intersection of HEP and ML.  To be best prepared for a career in this area, it is essential that researchers have a solid statistical foundation in addition to acquiring practical experience with applications.  Some programs are starting to develop along these lines, such as the \href{https://physics.mit.edu/academic-programs/graduate-students/psds-phd/}{PhD in Physics, Statistics, and Data Science} at MIT and the \href{https://data.berkeley.edu/decdse}{Designated Emphasis in Computational and Data Science and Engineering} at UC Berkeley.  However, training these researchers is not sufficient; we must also provide a career path for them within HEP.  ML is a catalyst for cross-cutting research and HEP ML specialists could help connect traditionally isolated areas of HEP. It is imperative that we take a broad perspective to recruit and retain a diverse cross-section of scientists. Capitalizing on industry expertise with cross-disciplinary collaboration will require community assessments of the ethics of such arrangements and new devices for external collaboration with HEP experiments.

\section*{\raisebox{-2.5ex}{\includegraphics[width=0.1\textwidth]{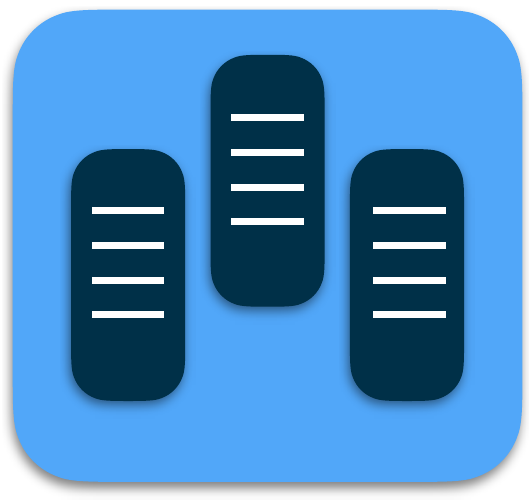}} 5 Storage and processing resource access}
\addcontentsline{toc}{section}{5 Storage and processing resource access}

The computing landscape has changed a great deal since the 2013 Snowmass Planning Exercise.  Some of the changes include dramatic changes in processors and widespread availability of heterogeneous nodes.  Also, there has been increased reliance on national supercomputer centers for the experimental program, and cloud computing has developed significantly.  We are just at the dawn of the Exascale era.  In the US, these computers all feature graphical processing units (GPUs) and generally have robust storage systems, but they are not ideally suited for much of the HEP workload.  Distributed computing in which multiple and far-flung computers and storage units are employed is very much the norm in HEP computing, so networking considerations are also essential for this topical group.  The Storage and Processing Resource Access topical group has focused its attention on the implications of these developments on research and development challenges that must be addressed in the next decade.  The group did not attempt to assess the required capacity of either computers or storage across the spectrum of HEP activities in that period.

This topical group divided its efforts into six areas: storage; processing; edge services; AI/ML hardware; analysis facilities; and networking.  There were some common issues that we will focus on before looking at specific areas.  High energy experiments are often international efforts.  Their computing resources can be globally distributed with
different nations providing compute and storage resources.  Thus, there is global flow of data among these resources.  CPU speeds have not been increasing for many years and increased processing capability comes from parallelism with multicore, wide vector or massively multithreaded hardware, with some computers applying all three.  In addition, processors designed for machine learning, FPGAs, and tensor processing units add to the heterogeneity of the current computing environment.  This leads to both an opportunity to increase the cost effectiveness of computing and the research and development challenge of making use of this diversity.

Four overarching themes, each of which appears in several sections of the report, have emerged:
\begin{enumerate}
    \item {\bf Efficiently exploit specialized compute architectures and systems}
    \item {\bf Invest in portable and reproducible software and computing solutions to allow exploitation of diverse facilities}
    \item {\bf Embrace disaggregation of systems and facilities}
    \item {\bf Extend common interfaces to diverse facilities}
\end{enumerate}

Exploiting specialized architectures will require the
allocation of dedicated facilities to specific processing steps in the HEP workflows, in particular for
``analysis facilities'' (discussed in topical group report Secs.~4.2 [Processing] and 4.5 [Analysis Facilities]); designing effective benchmarks to exploit AI/ML hardware
(report Sec.~4.3 [AI/ML Hardware]); improved network visibility and interaction (report Sec.~4.7 [Networking]); and enhancements to I/O
libraries such as lossy compression and custom delivery of data (report Sec.~4.4 [Storage]).

The need for investment in portable software libraries, abstractions and programming models is particularly ubiquitous and detailed in Secs. 4.2--6.  The last two sections discuss the need for software frameworks to enable reproducible HEP workflows.

Large HEP experiments cannot provide a single computing center for all their needs so must embrace the disaggregation.  This will require research on efficient distribution of workflows on heterogeneous facilities, including software abstraction to integrate accelerators, orchestration of network resources, exploiting computational storage and rack-level memory pools, and is detailed in Secs.~4.3, 4.4 and 4.7.

The fourth theme of extending common interfaces to diverse facilities is essential to enable use of those facilities by the many scientists who might make use of them.  It is also important for multiple experiments to embrace these interfaces to reduce development costs.  We also must not lose sight of the global context of our computing tasks.  The OSG Consortium and Worldwide LHC Computing Grid (WLCG) represent a successful style of distributed computing.  HEP must continue to encourage edge-service platforms on dedicated facilities as well as Cloud and HPC, and develop portable edge-services that are re-usable by other HEP projects and exploit commonality within HEP and other sciences (Sec.~4.6). These interfaces will
also need to extend into all aspects of HEP workflows including data management and optimising data
movement (Secs.~4.2, 4.4 and 4.7); as well as the deployment of compute resources for analysis
facilities (Sec.~4.5).

It will be challenging to train and retain a workforce that can deal with these research and development challenges.  For instance, there is currently no single programming model to deal with offloading tasks to computational accelerators such as GPUs and FPGAs.  Each of the major GPU vendors for upcoming exascale computers has its own approach.  For NVIDIA, it is \href{https://developer.nvidia.com/cuda-toolkit}{\texttt{CUDA}}; for AMD, \href{https://rocmdocs.amd.com/en/latest/Programming_Guides/HIP-GUIDE.html}{\texttt{HIP}}, and for Intel, \href{https://oneapi-src.github.io/DPCPP_Reference/}{\texttt{DPC++}} or \href{https://www.khronos.org/sycl/}{\texttt{Sycl}}.  In addition, OpenMP is developing better offload capabilities, and there is a movement to incorporate parallelism within standard languages such as \texttt{C++}.  The DOE is investing in the \href{https://alpaka.readthedocs.io/en/0.5.0/usage/intro.html}{\texttt{Alpaka}}, \href{https://github.com/kokkos/kokkos}{\texttt{Kokkos}}, and \href{https://github.com/LLNL/RAJA}{\texttt{Raja}} approaches to portability to simplify the task of writing code that a compiler can optimize for different architectures.  Expertise in dealing with security issues required for distributed computing is also rare.

In regards to processing, there are four recommendations.  
\begin{itemize}
    \item HPC facilities should revisit their resource access policies to allow more flexible allocations and job executions. This, coupled with new authentication and authorization models, will allow more HEP projects to benefit from the large computing facilities. 
    \item Investment in software development effort is key to maximize the efficient utilization of diverse processing resources. In particular, research and development of portable software solutions is critical for a sustainable software ecosystem in light of the evolving and increasingly diverse hardware architectures. 
    \item Research is needed to determine the tradeoff between dedicated HEP computing facilities and general-access computing facilities such as the HPC center, Grid and Cloud resources. 
    \item Infrastructure development will be needed to support better data management frameworks across different types of facilities.  
\end{itemize}

Hardware for AI/ML is quite diverse and changing rapidly.  Although there are many machine learning benchmark suites, a suite focused on HEP applications should be developed and used on current and new hardware in order to select the best hardware for our needs.  AI/ML hardware can considered to be either directly connected or via a network denoted ``as a service.''  %

Storage requirements for LHC experiments CMS and ATLAS are vast and present a major challenge given current and projected hardware budgets.  We anticipate that tape will continue as the archival medium.  Spinning and solid state disks will continue to co-exist, though the cost advantage for spinning disks continues to favor them for the bulk of the required capacity.  Research is underway on new data formats that are smaller or more efficient for modern processors.  \href{https://www.hdfgroup.org}{\texttt{HDF5}} may be useful at HPC centers with parallel file systems.  I/O frameworks will be needed to evaluate data-format performance (building on existing work conducted under HEP-CCE for example~\cite{compf4}).  Work is needed on both lossless and lossy data compression to reduce storage costs.  The separation of storage and processing presents a challenge to the network and one mitigation is to provide ``computational storage,'' in which some compute capability is embedded within the storage.  There are a number of technologies that need to be tracked as detailed in Sec. 4 of the topical group report.

Infrastructure and services will be needed that provide integrated data, software and computational resources to execute one or more elements of an analysis workflow. For example, the HL-LHC will present a daunting challenge to physics analysis teams.  To prepare for the that challenge IRIS-HEP is organizing an ``\href{https://iris-hep.org/projects/agc.html}{Analysis Grand Challenge}'' to test new concepts for and prototypes of analysis facilities.  The Analysis Grand Challenge workflow defines an analysis benchmark that could be easily re-implemented
and executed on any generic Analysis Facility and is designed to help to showcase to physicists how to use an
existing analysis facility at scale for their analysis.  Several contributed papers~\cite{compf4} discuss new concepts and features to be investigated.  Modern analysis tools such as machine learning, and fitting as a service run well on high performance computers and should be integrated with analysis facilities.  Security in distributed computing environments is a critical issue and work on federated identification needs to continue.  For example, NERSC is currently extending its authentication to allow users to login via credentials from their home institution.  This is enabled for most national laboratories.  Section 5 of the topical group report provides many more details.

Edge services operate at the interface between a data center and the wide area network, separated from the data center’s core services. This includes middleware that facilitates user access between
the data center and external systems (e.g., storage, databases, workflow managers). These services may be managed externally in partnership with the data center and federated across multiple data centers.  Containers can provide executable code for services in a portable manner, as they contain all the needed software and libraries.  Kubernetes is one example of system to create and distribute containers with very widespread industry usage, that will require further development to fully adopt into HEP workflows and resources.  We have already mentioned security and federated identity services as requiring additional R\&D.   We expect that HPC centers and HEP experimenters will make increasing use of edge services, so it will be necessary to follow technical developments in this area.

In a world of distributed storage and computing, the network is essential for getting work done.  Particle physics network traffic is expected to grow by a factor of ten by the end of this decade.  (For details, please see the recent \href{https://escholarship.org/uc/item/78j3c9v4}{ESNet HEP Network Requirements Review}.)  Four interrelated networking issues have been identified: network interaction optimization; resource orchestration and automation;  network and traffic visibility; and data movement optimization.  The first issue ensures that applications efficiently put packets on the network.  The second issue is one of coordinating schedule activities of storage systems, compute infrastructure and the network in order to reduce delays in a workflow.  The third and fourth capabilities are necessary for efficient orchestration since there must be a way to estimate the time to move data and the best way to route it.  These four areas must be worked on together.  The technology landscape changes quickly over time, and well-organized
collaborations staffed with knowledgeable experts can make effective use of current and future technologies,
whatever they may be. It is critical that particle physics make long-term investments in collaborations
between scientists and technologists so that cutting-edge networking technologies can be effectively used by
us. These collaborations must combine research, prototyping and production implementation,
as this is the only way that components and technologies can be effectively integrated into the scientific
enterprise as effective capabilities.  Prototypes offer powerful means of demonstrating new technologies and capabilities, allowing evaluation of cost, complexity, effort and maintainability. In all the identified areas, we suggest that there be work plans that clearly identify the steps and decision points from prototyping to production.  Each of the four areas is detailed in Sec.~7 of the topical group report.

\section*{\raisebox{-2.5ex}{\includegraphics[width=0.1\textwidth]{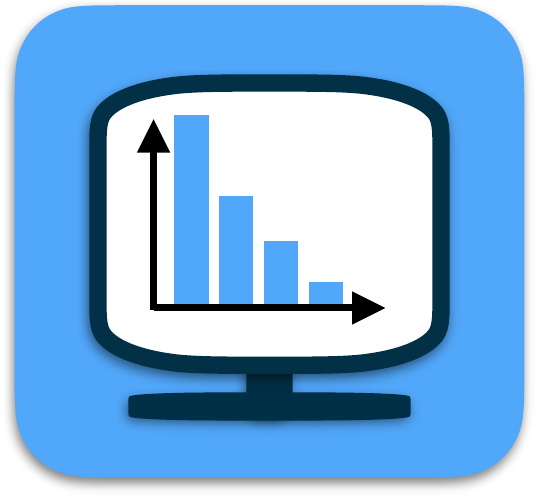}} 6 End user analysis}
\addcontentsline{toc}{section}{6 End user analysis}

End user analysis infrastructure is of fundamental importance to guarantee the fast turnaround in the process of running analysis code on full data sets with the frequency needed for timely delivery of physics results to be presented in conferences and published in scientific journals. Fast, easy, and user-friendly access to analysis level data formats is a primary premise in the collaborative process that allows analysis teams to assemble all elements of a physics measurement and iterate efficiently and quickly to produce a final result and go through internal and external peer review.

No analysis software functions entirely on its own, but in the context of the input data it consumes, the output data it produces, the other software it depends on, and the way it is configured or embedded in other code. We therefore talk about a ``software ecosystem.'' Packages in an ecosystem typically have common data interchange formats and similar programming language interfaces. There are two major ecosystems relevant to data analysis in HEP, based on ROOT and Python. The ROOT suite is a set of libraries that covers a broad range of needs, including I/O, event loop execution, histogramming, statistical analysis, and visualization. It is hosted by CERN with a leading contribution from the US to I/O. The libraries are written in \texttt{C++}, although the Python bindings (\href{https://root.cern/manual/python/}{\texttt{PyROOT}}) are well supported and expose essentially the full API.  There is also \href{https://github.com/scikit-hep/uproot5}{\texttt{UpROOT}} (see also \href{https://scikit-hep.org}{\texttt{Scikit-HEP}}) which is pure Python and does not require compiling ROOT.  Bindings for the R language are also currently supported. The tightly-bound nature of ROOT means that using alternative software for any particular functionality can be very difficult. Of particular interest is the planned evolution of the TTree into the n-tuple and nested tuple (\href{https://root.cern/doc/master/md_tree_ntuple_v7_doc_README.html}{\texttt{RNTuple}}) columnar data storage. This development will speed up analysis by improving the mapping to vectorized and parallel hardware and provide better memory control.
Python is a loose term for a set of tools, with Python as the primary language interface, introduced with the primary goal of enabling the use of software developed outside of HEP, including for machine learning. This ecosystem is still in development and has no single governance team. It tends to emphasize independent packages for different aspects of the analysis pipeline. Development teams in this ecosystem within HEP are typically small and feature junior personnel.  While there are a number of HEP developments within \texttt{Scikit-HEP}, it would be interesting to consider to what extent the HEP community can and should contribute to the broader scientific Python ecosystem.

Users interact with software libraries and packages through programming languages. These can be separated into general-purpose languages (GPLs) which can be used for any task, and domain-specific languages (DSLs) which provide a restricted set of higher-level primitives which simplify certain operations. The two most commonly-used general-purpose languages in HEP are \texttt{C++} and Python. Examples of domain-specific languages are the \href{https://root.cern.ch/doc/master/classTCut.html}{\texttt{TCut}} syntax used for applying selections and constructing new variables in ROOT, and the \texttt{kumac} language used to control the old FORTRAN-based \texttt{PAW} suite. DSLs have been historically easier to master than GPLs due to the limited range of constructs available. DSLs also have the advantage of providing high-level primitives which facilitate parallelization and acceleration. Python came to be adopted by users as a convenient language and its simultaneous adoption as the standard language in machine learning has driven massive bottom-up adoption of the language. However, since it can be slower than \texttt{C++} for some tasks, there is interest in exploring languages that can combine its expressiveness and ease-of-use with compilation to machine code; the most commonly-explored option is \texttt{Julia}.  There are also emerging paradigms within Python (and \texttt{Julia}) that use differentiable and probabilistic programming for optimal analysis.

Visualizations are increasingly being done via web browsers such as \href{https://github.com/jupyter}{\texttt{Jupyter}} notebooks or \href{https://root.cern.ch/js/}{\texttt{JSROOT}}, which follow common standards to offload rendering and interaction to the browser, resulting in a more uniform experience across platforms and reduced external dependencies.  This is the baseline for future ROOT graphics. As the dominant language in that environment is JavaScript, expertise will need to be maintained at some level. End-user analysis relies critically on visualization to give feedback to the physicist. Plots allow the user to quickly understand characteristics of the data but also to debug code and workflow problems. Although event displays are not typically used as part of an analysis workflow, they are still of interest to end users, and the same issues apply. Work is being done on experiment-agnostic event displays that render in browsers using JavaScript.

Data formats target different processing and analysis needs. Event data is encoded in a fixed but complex schema, describing individual events. Histograms summarize features extracted from event data. There are forms of summary data that cannot reasonably be expressed via histograms and they are presented in a tabular format. The configuration for running some software may be stored in a human-readable and -editable format, or in a binary format. File formats can describe both the overall container for data and the specific types of objects that can be stored. ROOT separates these fairly strictly, in that the ROOT file format can store essentially any \texttt{C++} object, and ROOT objects can be serialized to formats other than ROOT files (such as JSON or XML). ROOT provides a number of pre-defined data objects, such as tables (known as TTrees) and histograms, and multiple objects can be present in the same file. 

Once analysis code is written, it must be run on data. An analysis pipeline may involve multiple stages of data reduction and different codes, executing on very different platforms. Because the rapidity with which analysis workflows can complete is paramount, data processing architectures which are well-suited to managed production workflows may not match well on to analysis tasks. Users need to be able to scale code execution from a few events (for testing) to an experiment's full dataset. The former requires interactive response, while the latter may require distributed execution. Consequently, solutions that smoothly scale from small-scale interactive tests to full-data processing are desirable.  Interactive use historically has occurred via terminal sessions, while distributed execution has occurred on the grid, with complex issues of data locality and access, bookkeeping, job brokering, fair access, arising.
Particle physics problems are usually \textit{high-throughput}, not \textit{high-performance}, problems. They consist of a very large number of fairly lightweight computations which are essentially independent of each other, and so do not require computing resources to appear as a single, very powerful image (as on a traditional supercomputer). Traditional tightly-coupled supercomputer execution environments are therefore not typically needed for HEP analysis applications and in fact may be detrimental as they do not exploit the fine granularity of HEP problems. However, the increasing exploitation of coprocessors and accelerators (such as GPUs) in HEP code requires libraries that couple CPU and GPU execution on a single node. One line of research explores analysis facilities which would provide users with alternate mechanisms to access computing resources and enable new programming paradigms. \href{https://coffea-casa.readthedocs.io/en/latest/}{\texttt{Coffea-Casa}} is a prototype analysis facility, which provides services for ``low latency columnar analysis'', enabling rapid processing of data in a column-wise fashion. This provides an interactive experience and quick ``initial results'' while scaling to the full size of datasets. These services, based on the \href{https://www.dask.org}{\texttt{Dask}} parallelism library and \texttt{Jupyter} notebooks, aim to dramatically lower the time for analysis and provide an easily scalable and user-friendly computational environment that will simplify and accelerate the delivery of particle physics measurements.

In order to serve the long-term end-analysis needs of the community in a sustainable way, it is important to take action on a number of recommendations. Firstly, {\bf software work (especially with cross-experiment application) should receive stronger consideration for funding.} More cross-experiment/frontier computing physicist positions could be created. Funding agencies and frontiers need to work together to identify viable long-term funding patterns for this work.

More specifically:

\begin{itemize}

\item {\bf Develop both ROOT-based and Python-based analysis ecosystems.} The friendly competition between the two has already resulted in significant improvements for users. Maintaining interoperability between the two, e.g., data formats, should be a requirement.
\begin{itemize}
\item Critical elements of the ROOT evolution plans include the development of improved columnar data formats (evolution of TTree to the more-optimized RNtuple) and the multiple I/O implementations.
\item Analysis facilities for low latency columnar analysis in the context of the Python ecosystem will be essential for future experiments.
\end{itemize}
\item Effort should be put into {\bf developing user-friendly data provenance and metadata storage systems} that can be easily integrated into typical analysis tasks.
\item {\bf Scalable analysis models} should be developed to allow users to perform interactive tests and run over large datasets using a single interface.
\item {\bf Pipelines compatible with long-term preservation} should be built into the structure of analysis systems as the default.
\item {\bf Collaborative software} is an important element of the analysis software stack of an experiment, and includes documentation, messaging between users, discussion forums, software version control, bug tracking, and document workflow management. Host laboratories should provide a full stack of these services to their experiments, large and small.
\item {\bf Documentation and training} must be produced for analysis software understood as an ecosystem, not as a disconnected set of packages.
\end{itemize}

\section*{\raisebox{-2.5ex}{\includegraphics[width=0.1\textwidth]{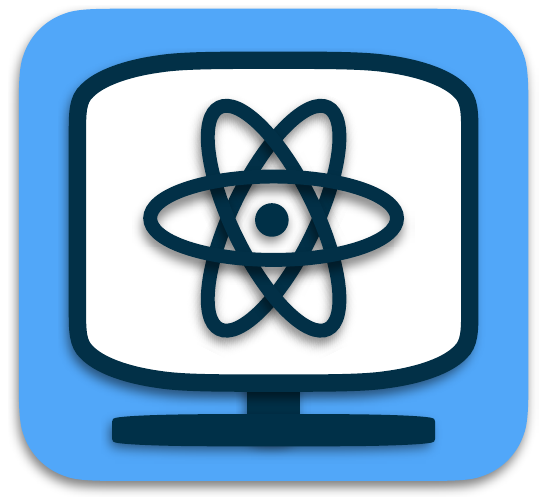}} 7 Quantum computing}
\addcontentsline{toc}{section}{7 Quantum computing}

Quantum computers (QCs) exploit quantum effects such as interference and entanglement to perform calculations. These devices are like classical computers in that they can be (re)programmed to perform many different calculations.  However, current QCs are relatively small and noise is a significant challenge.  This has naturally divided research into R\&D on near-term devices and R\&D for future \textit{fault tolerant} computers.  The setup of near term QCs is similar to a small HEP experiment at a user facility in terms of the steps to initiate and process a measurement.  Similar to other HEP experiments, we are striving to discover new insight with these experiments.  The research and experience with near term computers may also inform R\&D for the fault tolerant era.  While quantum computing is closely related to other areas of quantum information science (theory, sensing, and networking), we focus solely on the use of reprogrammable quantum hardware (`quantum computers') in this section.  Rapid progress in hardware and software over the last years has enabled a rapidly growing research area of software for QCs.  Open source tools exist for writing algorithms to manipulate quantum computers and these algorithms can be deployed on quantum hardware similarly to a classical batch system.  These significant advances have motivated the dedicated topical group on QC for HEP.

Even relatively small QCs can outperform classical computers for certain questions.  This is because QCs can represent an exponentially large Hilbert space with polynomially many qubits.  However, not all exponentially hard problems can be efficiently solved on a quantum computer.  It is also possible that fault tolerance will be required for going beyond classical computers.  Nonetheless, it is necessary for HEP scientists to participate in near-term computing efforts (called the NISQ era~\cite{Preskill_2018}) in order to develop motivational examples, to grow a QC for HEP workforce, and to build critical infrastructure for the future.  While the timescales are not clear, it is possible that fault tolerance will be achieved for the next Snowmass process in about a decade.  It is clear that QC technology will continue to improve at a rapid pace over the next years.

A number of benchmark problems have emerged which have focused the QC for HEP community and which are particularly promising.  These include lattice gauge theory, event generation, and data analysis.  While classical lattice gauge theory is a well-established tool to solve a number of questions that cannot be addressed with perturbation theory, there are significant limitations (the \textit{sign problem}) for answering all non-perturbative questions~\cite{Bauer:2022hpo}.  Lattice gauge theory on a quantum computer may be able to solve these challenges.  Hybrid methods that combine classical and quantum computations are also promising, especially with near-term devices. Monte Carlo event generators are also able to achieve excellent precision for many observable, but there are some regimes where quantum effects are large and hard to model with classical methods.  The potential for classical data analysis with quantum computers is less clear because of the significant overhead to transform the data itself into a quantum representation.  Nonetheless, there are cases where the best quantum algorithms are at least polynomially faster than their corresponding classical counterpart and other cases still where early empirical results are promising.  Gains may be larger for data from quantum sensors that are inherently quantum.  Continued investment in identifying benchmark problems is crucial, also to engage the non-HEP QC community.

A significant difference between quantum and classical computers is that there is no consensus on the hardware underlying quantum computations.  Candidate technologies include superconducting qubits, trapped ions, and photonic systems.  Some systems are publicly available and others require specific agreements with particular laboratories or companies.  Access to quantum computers is a challenge for researchers and this barrier could be eliminated with coordination across the HEP program.  There are also exciting opportunities for co-design whereby quantum computers could be constructed that are tailored for HEP applications.  This important research area will require collaboration between HEP QC experts and QC hardware researchers (likely not in HEP).  More generally, the connection between QC and HEP is not one way as the HEP community has expertise that could benefit QC generally.  Taking advantage of partnerships between HEP and non-HEP entities (including national QIS centers) is an effective strategy for mutually beneficial research. 

There are also a diverse set of software packages for specifying quantum gate sets and for compiling them to particular quantum devices.  It may be beneficial for HEP to directly contribute to these packages to ensure that the evolving HEP needs are met in addition to providing a service to the broader community.  Additionally or alternatively, it may be beneficial to create a community-wide codebase for HEP applications similar to \href{https://qiskit.org/documentation/nature/}{\texttt{Qiskit Nature}}.

As with many areas of software and computing, researchers with QC experience are in high demand.  A growing number of Computer Science and Physics Departments are offering courses in QC, and these could be a component of the training path for a researcher at the intersection of HEP and QC. It is also vital that we forge career paths for these researchers, who often do not neatly fit into one of the existing frontiers.  There is also an opportunity to build connections with industry, both for the science and for the career development of junior researchers.

There is also a strong synergy between QIS for HEP and QIS more generally.  For example, many HEP scientists are members and leaders of the National QIS Research Centers and one of the centers (Superconducting Quantum Materials and Systems Center (SQMS)) is hosted at Fermilab.  SQMS is leveraging decades of investment in superconducting technology in HEP to understand the physics of decoherence, and to build a quantum information processing platform based on superconducting radio-frequency (SRF) cavities. These processors may be particularly well suited to quantum simulation problems of interest inside and outside of HEP.  %

\section*{\raisebox{-2.5ex}{\includegraphics[width=0.1\textwidth]{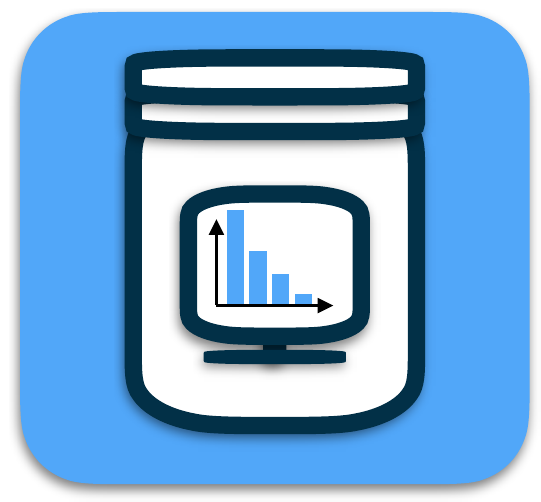}} 8 Reinterpretation and long-term preservation}
\addcontentsline{toc}{section}{8 Reinterpretation and long-term preservation}

The topical group devoted to Reinterpretation of Long-Term Preservation of Data and Code has considered what needs to be done to achieve these goals and the barriers to achieving them.  In addition, it is argued that a facility should be set up to preserve data from Cosmic Frontier observations and simulations.  This is just one example of the need for facilities for long-term data preservation, as discussed below.

Preserving data, simulations, analyses and the codes needed to do the analyses is a practice that can increase scientific output from the investments made in the original inquiry.  Most scientists see that this is a benefit\footnote{Although there is some disagreement about when and how data products and software should be made public.} to the wider community and, in principle, support this activity.  However, to actually plan for and carry out such preservation requires resources and effort from scientists who many not have a strong incentive to do this work.  The large, long-running experiments such a those at the LHC are in a better position to plan for long-term preservation than the smaller experiments.  This is due to CERN support for data preservation and the fact that larger experiments tend to have more personnel devoted to S\&C.

We recommend:
\begin{enumerate}
    \item Ensuring that all running and in-preparation experiments have a strategy and resources 
for the long term preservation of data and analysis capabilities, extending beyond the lifetime of the
 experiment.
    \item Investing in shared cyberinfrastructure to preserve these data and support a comprehensive analysis from
various experiments and surveys---both active and completed---in order to realize their full scientific
impact. The infrastructure should support the requisite theoretical inputs and computational requirements
for analysis as well as metadata and APIs to track provenance and incentivize participation.
   \item As mentioned above, as a specific example, there should be a data archive center to preserve Cosmic Frontier datasets and simulations, and facilitate their joint analysis across different computing centers.
\end{enumerate}

We note that preservation needs are not restricted to experimental data and analyses.  As an example, for many years, the lattice QCD community has had a tradition of sharing the expensive-to-produce gauge configurations that can be used for many physics analyses.  NERSC hosts the \href{https://qcd.nersc.gov}{Gauge Connection} which is one facility for such sharing.  The Gauge Connection has been described as their first web portal.  Another effort started 20 years ago is the \href{https://hpc.desy.de/ildg/}{International Lattice Data Grid}.  However, these efforts are restricted to one type of data.  Several groups have also made their simulation codes available, and there has been SciDAC funding to support \href{https://www.usqcd.org/software.html}{community software} that has been widely used.

Ideally, publicly released data should follow the data management principles of  being Findable, Accessible, Interoperable, and Reuseable (FAIR) as
described in Ref.~\cite{FAIRREF}.  Achieving this goal requires attention and resources throughout the end-to-end lifecycle of data generation, processing, analysis, preservation, and distribution.  Reference~\cite{Bailey:2022tdz} defines important concepts such as data, derived data, and analysis data products, which distinguish between the ``raw'' data coming from an experiment or simulation, to data reduced in some form, such as particle momenta and identities, to further products of analysis such a summary plots, etc.  Data preservation is the procedures, practices, and standards of ensuring the long-term (i.e., decades
beyond the end of an experiment) preservation, accessibility, and usability of data and derived data
from experiments.  Data analysis preservation is similar but its goal is to be able to repeat the analysis from the preserved data and reproduce the analysis data products.  Two addition concepts are reinterpretation, i.e., any type of new, alternative or updated interpretation of an experimental analysis
or result, including the combination in, e.g., global fits or global averages, recasting, which is reproducing the analysis logic in a simulation, considering a different physical process
with a different phase space distribution, which might have different efficiencies and acceptances than
the originally hypothesised model.

Data preservation requires the long term curation, storage and distribution of the three data types defined above.  \href{http://opendata.cern.ch}{CERN's Open Data Portal} provides the infrastructure and services for CERN experiments, but there is no equivalent (in the US) for smaller experiments or theoretical data.  For astronomical data, NSF and NASA provide several archival storage facilities, but DOE does not provide such infrastructure for DOE-funded Cosmic Frontier experiments.  Coordination with national supercomputing centers would be welcome.

Preservation of analysis data presents its own challenges as there may be many formats.  \href{https://www.hepdata.net}{HEPData}, run by Durham University and CERN contains data from nearly 10,000 publications, with over 100,000 tables of data.  CERN provides the \href{https://www.zenodo.org}{Zenodo} site that is capable of sharing larger data sets and serves a wider community than high energy physics.  With such infrastructure, the challenge is to get scientists to routinely archive their data as part of the publication process.

Preserving analysis software and logic present their own challenges; however, services such a GitHub, Bitbucket, and GitLab help to make it easier to distribute open source analysis code.  Containerization can help to mitigate the dangers of operating systems or libraries changing in such a way that the code no longer works as it should.  Machine learning models create a challenge in that the underlying software and the format of its model may change in an incompatible way.  On the other hand, Jupyter notebooks are becoming very popular and their nonproprietary nature and ability to combine text, graphics, and code help to preserve the analysis logic.  In addition. \href{https://github.com/onnx/onnx}{onnx} provides an open standard for machine learning interoperability.

\section*{9 Personnel and Training}
\addcontentsline{toc}{section}{9 Personnel and Training}

Two critical topics that emerged for all of the topical groups are (i) educating researchers to apply and develop advanced software and computing skills and (ii) ensuring there is a career trajectory inside HEP for computational researchers.  The first topic is of great interest to other Frontiers, and also considered by the Community Engagement Frontier (CEF).  The CEF also has a topical group on Career Pipeline and Development.

The current standard for S\&C training is project-specific on-the-job training. These training activities can be very effective, especially with formalized curricula and documentation.  However, these activities are often inaccessible beyond a particular experiment or other organization and due to limited person-power, these events often do not cover as deep or as broad as is needed to be maximally effective with S\&C.  These challenges can be solved with common training activities with persistent and accessible documentation.  Some aspects of S\&C training are not HEP-specific and we can turn to general computing and data science training materials (some of which have been led by HEP researchers).  One example of this where HEP researchers have engaged larger efforts is through the \href{https://software-carpentry.org}{The Carpentries}. Other aspects of S\&C are HEP specific and are best taught in the HEP context.  In these instances, it is essential that we prepare training activities and widely-available documentation.  A success story in this area related to Snowmass is the recent DOE funding opportunity \textit{Traineeship in Computational High Energy Physics} (\href{https://science.osti.gov/hep/Funding-Opportunities/-/media/grants/pdf/foas/2022/SC_FOA_0002743.pdf}{DE-FOA-0002743}), which cites as motivation the 2013 Snowmass Computational Frontier report, in particular the one on Software Development, Personnel, and Training~\cite{Brown:2013hwa}.  A number of other training programs have been hosted by DOE- and NSF-funded initiatives as well as by the HSF and other organizations.

Career trajectories in HEP S\&C can take a number of forms.  Early exposure to S\&C topics can occur in undergraduate curricula.  We strongly support DOE and NSF fellowships for graduate students who want to specialize in or have a major focus on S\&C topics.  Beyond graduate school, paths should exist for physicists who are experts in S\&C as for physicists who dedicate their career to S\&C as a research software engineer (RSE) or equivalent.  In both cases, paths towards positions with long-term stability are essential.

\section*{10 Diversity and Climate within Computing in HEP}
\addcontentsline{toc}{section}{10 Diversity and Climate within Computing in HEP}

The percentage of Underrepresented Minorities (URMs) obtaining a Ph.D. in physics in the US is $<5\%$ (Hispanics) and $<2\%$ (Blacks), while the percentage of women is about 20\%~\cite{apsstats}. In computer science the situation is worse for Hispanics and Blacks, with approximately 1.5\% of Ph.D. degrees earned by the former, 1\% by the latter, while the percentage for women is about the same in computing and physics~\cite{crastats}. Representation of Hispanics and Blacks in both physics and computing stand in stark contrast with their numbers as a fraction of the US population, 19\% and 14\%, respectively. While gender and race disparities in physics and computing often receive the most attention, identity is multifaceted and individuals face adversity in the field based on their sexual orientation, religion, disabilities, immigration status, country of origin, and more.
Interestingly, the situation was not always so dire for women in the case of computer science. In 1970, the  percentage of women majoring in computer science was 12\%, increasing dramatically to a 37\% by the mid eighties, only to plummet and flatten out to a 17\% by 2010~\cite{nprarticle, HW:2021}.
Conversely, 7\% of the Ph.D.s in physics were awarded to women in 1970, increasing linearly to eventually flatten to a 20\% in the early 2000's~\cite{nprarticle}. There are several compounding factors that contribute to persistent diversity gaps in computer science including stereotype threat, implicit and explicit bias in higher education and society at large, isolation and lack of community, and more, as discussed in Refs.~\cite{HW:2021,MMC:2021,EKA:2019}.
As compared to the overall participation within HEP, the participation of women in HEP S\&C activities may be even lower. This would not be surprising, given that the prevailing culture within computing in HEP mirrors that of the computing community within society at large, creating climate-related barriers that undermine gender and race equality. 

Another challenge the S\&C community faces is related to the lack of recognition of the physics content of computing work within HEP, unlike the case of contributions to instrumentation, which are better rewarded career-wise. As the intersection of two highly
specialized topics, there is an especially high level of knowledge and skills required to contribute successfully to computing in HEP. To get to this level, individuals usually must seek expertise outside of traditional HEP education and training,
which typically does not cover computing topics in detail. This poses a discouraging burden, especially for those without connections to such expertise or without the resources to take on such additional work. It also creates in the physicist the perception that computing in HEP is an uphill and separate career path with different career success metrics and requiring a different set of skills. All these elements compound to create barriers which discourage participation, making recruitment and retention challenging among young physicists who are building their expertise on the computing aspects of HEP. Substantially expanded support for S\&C training and for software development as a viable career in the HEP ecosystem would alleviate this burden. Support for strong recruitment efforts and dedicated mentoring programs would help increase and nurture the talent pool. A larger and healthier HEP computing community can be more welcoming and accessible to participants from all backgrounds.

As a first step, the US HEP S\&C community should establish a study group to explore the challenges in the area of diversity, inclusion and climate that are particular to the computing sub-field within HEP. This group would be charged with the task to identify these challenges and produce recommendations to address them. Any effort to characterize and address diversity and inclusion issues in HEP and computing must be multifaceted and intersectional.

\section*{11 Recommendations}
\addcontentsline{toc}{section}{11 Recommendations}
\label{sec:recs}

Before expanding on the recommendations from the Executive Summary, we briefly revisit some important aspects of the Computational Frontier.  The Snowmass process is mostly about the long-term, post 2035 period, with next- (or in some cases, next-to-next-) generation projects across frontiers. However, the near-term program, especially from the point of view of S\&C, before 2035 is not completely determined, and we must make a strong case for it.  We have focused on the upcoming 10-15 years, where we need to secure funding to bridge the resource gap of near-future programs and to account for the rapidly evolving technologies that make predictions beyond 2035 inaccurate.  Despite this focus, we must also be cognizant of the near-term program required to address the needs of feasibility studies and R\&D for future facilities.  To complicate matters, computing technology and software paradigms change on a much faster scale than the life of particle physics experiments.  Additionally, many S\&C efforts are neither funded nor managed as projects, unlike facilities and experimental devices.

One of the purposes of the Snowmass reports is to inform the Particle Physics Project Prioritization Panel (P5).  In the 2014 report~\cite{HEPAPSubcommittee:2014bsm}, S\&C topics were listed in recommendation 29 (out of 29):

\begin{quote}
\textit{Strengthen the global cooperation among laboratories and universities to address computing and scientific software needs, and provide efficient training in next-generation hardware and data-science software relevant to particle physics. Investigate models for the development and maintenance of major software within and across research areas, including long-term data and software preservation.}
\end{quote}

The 2014 P5 report also has two paragraphs about S\&C in the context of enabling the HEP science drivers.  These paragraphs note that HEP played a leading role in high-throughput distributed/grid computing, online data processing, high performance computing and networking, large-scale storage, and the world wide web.  We have successfully managed software development and operations on a global scale and computing in HEP continues to evolve based on needs and opportunities.  Noted at the time were a variety of areas including (i) high performance computing and novel algorithms for realistic beams, (ii) the volume and complexity of physics data from LHC experiments stressing computing infrastructure and expertise (iii), cosmology programs extending data needs as vast new surveys and high-throughput instruments were coming online, and (iv) theory computations increasing in importance as higher fidelity modeling was required.
 
In 2022, S\&C continues to be an enabler of the HEP science drivers. Computing is essential to all experiments and many theoretical studies.  Data volumes, detector complexity, and precision required in calculations and simulation will continue to grow in near- and far-future experiments and surveys.  The size and complexity of the software and computing effort has grown to be commensurate with that of the experimental instruments. Projects may also need software-detector co-design to optimize physics performance at a minimum detector and computing cost. S\&C has evolved to be an increasingly collaborative and global effort across projects, frontiers, and fields, thus work should be coordinated with worldwide partners.
 
 At the same time, S\&C technologies are changing the face of HEP. We have solidified the trend towards computing hardware heterogeneity and specialization, and increased use of high-performance computing facilities.  AI/ML was neither on the horizon of the 2013 Snowmass process nor in the text of the 2014 P5 report, but it is now an established technology widespread in every HEP area.  Quantum computing is entering the stage as the emerging technology with potential impact on quantum many-body systems, event generators, data analysis, etc.
 
 With all of these considerations, we have focused on a single key recommendation followed by four other key recommendation areas.  We now expand on each of these topics.

\subsection*{Coordinating Panel for Software and Computing (CPSC)}

The CPSC draws inspiration from the analogous panel in instrumentation.  The Coordinating Panel for Advanced Detectors (CPAD), sponsored by APS' Division of Particles and Fields (DPF), \textit{seeks to promote,
coordinate and assist in the research and development of instrumentation
and detectors for high energy physics experiments} (\url{https://cpad-dpf.org}). CPAD was formed around the time of the last Snowmass and serves a number of purposes in the HEP instrumentation community. It is led by an executive committee of about a dozen researchers from universities and national laboratories.  Additionally, CPAD has a number of selection committees for awards targeting junior and senior researchers as well as for advising funding agencies.  Furthermore, the panel organizes an annual workshop and helps with coordinating various reports (such as the 2020 DOE Basic Research Needs study).  For computing in HEP, we envision a CPSC with similar properties and mandate, although the exact details would need to adjust to the different characteristics and requirements of the S\&C community. For information, we include a quote from the CPAD mandate, which is recorded in the \href{https://engage.aps.org/dpf/governance/bylaws}{DPF bylaws}:

\begin{quote}
\textit{The Coordinating Panel for Advanced Detectors (CPAD) shall consist of a Chair, as well as a Vice-Chair and six members appointed by the DPF Chair-Elect following a call for nominations and approval by the DPF Executive Committee prior to appointment by the Chair-Elect. Panel members shall serve staggered, two-year terms but should not serve more than two consecutive terms. The newly appointed member of the CPAD chair-line shall serve first as Vice-Chair, succeeding to the Chair position in the second year and serving for one year. The CPAD shall have the responsibility to promote excellence in the research and development of instrumentation and detectors to support the national program of particle physics in a global context through the organization of the annual topical meeting on detector research and development; the nomination and selection of the annual DPF Instrumentation Awards and the Graduate Instrumentation Research Award; the promotion of educational programs to further the understanding of detectors and their instrumentation; the organization of multidisciplinary workshops; and the development of new activities consistent with its mission.}
\end{quote}

We recommend that DPF append their bylaws to include a similar statement for a CPSC.  Central to the CPSC's mission will be to liaise with the funding agencies to ensure that the S\&C needs are being met for near-term projects, as well R\&D targeting longer-term programs.  Additional tasks such as workshops on relevant topics and HEP computing awards would be part of the remit.  The panel could also help promote standards and best practices such as publishing and citing software.  To be most effective, the CPSC should have explicit connections to existing organizations, such as experiments and collaborations, the HSF, and DOE/NSF funded centers and institutes, which could be achieved through liaison positions and/or panel membership.  On the other hand, the CPSC should not promote the work of any one research group and it should try to remove barriers to entry (for individuals and groups entering S\&C) to the extent possible.  Early career researchers should be represented or otherwise encouraged to participate. The CPSC will need a particularly broad membership because it will need to cover the needs of both experiment and theory. The CPSC should not provide funding for projects.

\subsection*{Long-term Development, Maintenance, and User Support of Software}

In addition to existing grants for short term R\&D, we need mechanisms for longer term S\&C support.  This includes personnel to work on long-term code development, maintenance, and user support as well as hardware, such as through computing allocations (storage and computing power), that can support multiyear/long-term software projects.  Many projects in need of long-term development are also applicable across multiple experiments/projects and so support is harder to identify with existing structures (such as experimental operations programs), but the reward is also large.  In a recent survey of the community, a majority of the respondents believed that software maintenance is underfunded~\cite{Agarwal:2022gno}.  Some, but not all, work in this area could be covered through an expansion of and coordination between experimental operations programs.

One software package that came up in almost every one of the Computational Frontier events organized as part of the Snowmass process is \texttt{Geant4}. This is an essential tool that large and small experiments across frontiers use to develop future detectors and for modeling existing instruments. At the moment, most of the US support for \texttt{Geant4} developers comes through large experiments in the energy and neutrino frontiers.  The challenge is that this does not easily allow for longer-term development independent of a specific experiment's acute needs and it also makes it difficult for smaller experiments (that often cannot support \texttt{Geant4} developers) to influence the future of the program.

A related topic is the long term development of event generators.  These tools lie somewhere between theory and experiment, making their support even more complicated.  The physics development of event generators requires work on theoretical models, but the packaging, maintenance, and user support of these critical software tools is computational work that may not be valued by the theory community. There are also many common physics and computational challenges associated with event generators that could be served well by common efforts undertaken by a diverse and cross-cutting collaboration, similar to MCNet (\url{https://www.montecarlonet.org}) in Europe, which could bring together the U.S. event generator community, and facilitate the sharing of resources and ideas.  


There are also a number of experimental tools that are used by multiple experiments where no one experiment is responsible for their long-term development, maintenance, and user support.  Examples of this include the reconstruction frameworks (e.g., \texttt{LArSoft} [\url{https://larsoft.org}] and \texttt{ACTS} [\url{https://acts.readthedocs.io}]), data processing frameworks (e.g., \texttt{Gaudi} [\url{https://gaudi-framework.readthedocs.io}]), and production systems (e.g., \texttt{Rucio} [\url{https://rucio-ui.cern.ch}]).  There is also the ubiquitous data analysis framework \texttt{ROOT} (\url{https://root.cern.ch}), and there is now a growing reliance on other community tools from the scientific \texttt{Python} ecosystem like \texttt{Numpy} (\url{https://numpy.org}) and \texttt{SciPy} (\url{https://scipy.org}).  Given the reliance of HEP science on these general packages, it may be a wise investment for HEP to contribute to their long-term development, maintenance, and user support.

Another critical area is the modernization, maintenance, and user support of S\&C for legacy data.  Running experiments and surveys have infrastructure to support the interoperability of data products, mostly for internal consumption, but this varies by subfield.  However, the data from HEP experiments often have utility long after the instruments are no longer taking data.  Empowering (re)analysis of legacy data is not a one time cost and requires modest, long-term support beyond the lifetime of an experiment\footnote{Legacy data preparation can also begin with Open Data initiatives before the end of an experiment.}.  The physics benefit could be significant, especially as new insight is developed and as future experiments are planned.

\subsection*{Cross-cutting Software and Computing R\&D}

Many HEP software and computing tools are broadly useful, while researchers developing these tools tend to be supported or otherwise directly connected to a single project in a particular frontier.  However, many computational methods and tools are broadly applicable across HEP and computational HEP as a discipline can catalyze cross-talk between HEP subdomains.

Enhancing (in contrast to maintaining) the tools from the previous section provides a class of important examples of an impactful investment in cross-cutting R\&D.  Another class of examples is advanced statistical methods, including machine learning, that can be applied to multiple HEP subdomains.  Like event generators, R\&D in this area is also between theory and experiment and often does not have a natural mechanism for continual support.  There is an urgent need to construct support mechanisms for cross-cutting Data Science within HEP, where researchers can develop, adapt, and deploy new methods to make the most of HEP data and simulations.  Deploying state-of-the-art methods may also require innovative hardware solutions that often transcend existing project boundaries.  Institutions like the DOE Center for Computational Excellence (HEP-CCE), the NSF Institute for Research and Innovation in Software for HEP (IRIS-HEP), and the NSF AI Institutes (IAFAI [\url{https://iaifi.org}] and A3D3 [\url{https://a3d3.ai}]) are great examples of interdisciplinary R\&D in this area and these should be continued and expanded.  Support at smaller scales (e.g., through individual PIs) is also important for a sustained effort.  In order to make the most of cross-cutting investments, there should be a plan to connect proof-of-principle research to applications. 

Enabling cross-cutting S\&C R\&D requires access to computational resources that are accessible.  A key challenge for cross-cutting R\&D teams is that they may not have access to computing allocations or common computing environments.  This should be addressed to ensure the success of research in this area.

Cross-cutting S\&C innovation can lead to significant jumps in performance, to enable science that was not possible before or to deliver physics results with significantly fewer resources than before.  This may lead to new discoveries, shorten the time from producing publication quality results, and reduce our carbon footprint.

Developments in this area (especially in the areas of machine learning and quantum computing) may also have broader implications beyond HEP, where HEP researchers may be able to make contributions with strategic and economic importance.


\subsection*{Heterogeneous Computing}

A growing fraction of HEP computing will be provided by hardware accelerators and/or distributed CPU machines from diverse sources (university cluster, HPC, grid, commercial cloud, etc.). We have an opportunity to make the best use of these resources by investing in personnel for code re-engineering and adaptation.  These personnel could be HEP scientists who are trained in code portability and/or software engineers.  In either case, a unique set of skills are required that are often not part of the standard HEP S\&C background.

Fortunately, the HEP community has some collective experience with porting software to be compatible with HPC resources, including with hardware accelerators.  In addition to many efforts within experiments and surveys, large and small, there are a number of dedicated programs such as the DOE Exascale Computing Project (ECP), HEP-CCE, Scientific Discovery through Advanced Computing (SciDAC), Computational HEP (CompHEP) more generally, and the NSF IRIS-HEP.  In addition to these programs, there are others like the NERSC Exascale Science Applications Program (NESAP) that have supported projects with personnel and early computing allocations.  Some of these projects are supported directly or in part by HEP (through the DOE Office of Science HEP program, the CompHEP subprogram and the NSF Elementary Particle Physics programs) and some efforts are supported through the DOE Advanced Scientific Computing and the NSF Office of Advanced Cyberinfrastructure programs.  Continuing and expanding programs across the HEP community and for projects of various scales will be critical in the future.

\subsection*{Training, Career Development, and EDI}

First and foremost, research in S\&C should be recognized as physics research, analogously to R\&D in instrumentation.  Correspondingly, there should be a clear career path in Computational HEP from undergraduate courses and research experiences to graduate school programs all the way to faculty and (physics) staff positions. Additionally, researchers who do not make Computational HEP their focus will still need training in S\&C.  Acquiring these skills should be part of our undergraduate and graduate educations in both experiment and theory.  Summer schools and other limited-scope training events can help augment the skill sets, but do not replace coursework. 

We hope that elevating the physics recognition for S\&C within HEP, will lead to the creation of tenure-track faculty and staff positions.  While funding agencies cannot directly create faculty lines, they could catalyze the creation of faculty positions through joint or bridged positions\footnote{Researchers are hired as regular faculty, but typically 50\% of their time is covered by a national laboratory through tenure.} with national laboratories.  This could be an effective tool for encouraging new positions, but we should be aware of the additional demands on the researchers for serving two departments/institutions instead of one.  The creation of cross-cutting tenure-track staff positions at national laboratories is another effective strategy.

Second, and equally important, we need to eliminate barriers for researchers with diverse backgrounds to work on S\&C and we need to create an environment that will retain this workforce.  To this end, the CPSC should establish a committee to study the status and challenges related to diversity, equity, and inclusion within S\&C in HEP.  There will be some challenges that are common with HEP more generally, but there may also be properties of S\&C that require specific attention.  The committee should present a list of actionable items to the CPSC leadership for implementation in coordination with the broader S\&C community. %

\section*{12 Conclusions and Outlook}
\label{sec:outlook}
\addcontentsline{toc}{section}{12 Conclusions and Outlook}

Software and computing technologies are evolving rapidly and it is hard to imagine what the state of the art will be at the time of the next Snowmass in about a decade. However, it is clear that the trend towards more heterogeneity and a diversity of software tools will continue, triggering the need for a HEP workforce with a correspondingly diverse set of skills and backgrounds.

S\&C traverses all areas of HEP.  While we have organized the Computational Frontier along cross-cutting topics, there are strong connections to all other Snowmass frontiers.  Experimental projects, large and small, in the Energy, Neutrino, Rare Processes and Precision, and Cosmic Frontiers will continue to rely on S\&C to deliver the best physics output from our instruments at a variety of latencies and bandwidths.  Many areas of HEP simulations and theory more generally also have stringent computational requirements.  There are also strong and growing connections to the other Snowmass Frontiers, including the Accelerator Frontier (e.g., accelerator control and modeling), the Instrumentation Frontier (e.g., experimental design), the Underground Facilities Frontier (e.g., quantum computing), and the Community Engagement Frontier (e.g., training and industry connections). 

It is our hope that the establishment of a Coordinating Panel for Software and Computing (CPSC) will serve the community on timescales commensurate with the rapid evolution of S\&C technologies.  An important function of this panel is to provide guidance to the funding agencies with input from the domestic and international HEP communities.  %
Advances in S\&C have been essential to progress in both experiment and theory.  Working together as a unified community, we can ensure this continues in the future.

\bibliographystyle{jhep}
\bibliography{refs.bib}


\end{document}